\documentclass[english,prl,final,twocolumn, superscriptaddress,lengthcheck]{revtex4-1}
\usepackage[T1]{fontenc}
\usepackage[latin9]{inputenc}
\usepackage{color}
\usepackage{babel}
\usepackage{textcomp}
\usepackage{amsmath}
\usepackage{amssymb}
\usepackage{graphicx}
\usepackage{esint}
\usepackage[unicode=true,pdfusetitle,
 bookmarks=true,bookmarksnumbered=false,bookmarksopen=false,
 breaklinks=false,pdfborder={0 0 1},backref=false,colorlinks=true]
 {hyperref}

\makeatletter
\@ifundefined{definecolor}{color}{}
\@ifundefined{definecolor}{\@ifundefined{definecolor}
 {\usepackage{color}}{}
}{}

\@ifundefined{definecolor}{\@ifundefined{definecolor}
 {\@ifundefined{definecolor}
 {\usepackage{color}}{}
}{}
}{}
\usepackage{babel}
\usepackage{longtable}

\@ifundefined{textcolor}{}{%
 \definecolor{BLACK}{gray}{0}
 \definecolor{WHITE}{gray}{1}
 \definecolor{RED}{rgb}{1,0,0}
 \definecolor{GREEN}{rgb}{0,1,0}
 \definecolor{BLUE}{rgb}{0,0,1}
 \definecolor{CYAN}{cmyk}{1,0,0,0}
 \definecolor{MAGENTA}{cmyk}{0,1,0,0}
 \definecolor{YELLOW}{cmyk}{0,0,1,0}
 }

\usepackage{epsfig}

\usepackage{amsfonts}\usepackage{float}

\makeatother

\begin{document}

\title{Precision limits of tissue microstructure characterization by Magnetic
Resonance Imaging}

\author{Analia Zwick}

\affiliation{Centro Atómico Bariloche, CONICET, CNEA, S. C. de Bariloche, Argentina.}

\affiliation{Departamento de Física Médica, Instituto de Nanociencia y Nanotecnologia,
CONICET, CNEA, S. C. de Bariloche, Argentina.}

\author{Dieter Suter}

\affiliation{Fakultät Physik, Technische Universität Dortmund, D-44221, Dortmund,
Germany}

\author{Gershon Kurizki}

\affiliation{Chemical Physics Department, Weizmann Institute of Science, Rehovot,
Israel}

\author{Gonzalo A. Álvarez}
\email{gonzalo.alvarez@cab.cnea.gov.ar}

\selectlanguage{english}%

\affiliation{Centro Atómico Bariloche, CONICET, CNEA, S. C. de Bariloche, Argentina.}

\affiliation{Departamento de Física Médica, Instituto de Nanociencia y Nanotecnologia,
CONICET, CNEA, S. C. de Bariloche, Argentina.}

\affiliation{Instituto Balseiro, CNEA, Universidad Nacional de Cuyo, S. C. de
Bariloche, Argentina.}
\begin{abstract}
Characterization of microstructures in live tissues is one of the
keys to diagnosing early stages of pathology and understanding disease
mechanisms. However, the extraction of reliable information on biomarkers
based on microstructure details is still a challenge, as the size
of features that can be resolved with non-invasive Magnetic Resonance
Imaging (MRI) is orders of magnitude larger than the relevant structures.
Here we derive from quantum information theory the ultimate precision
limits for obtaining such details by MRI probing of water-molecule
diffusion. We show that already available MRI pulse sequences can
be optimized to attain the ultimate precision limits by choosing control
parameters that are uniquely determined by the expected size, the
diffusion coefficient and the spin relaxation time $T_{2}$. By attaining
the ultimate precision limit per measurement, the number of measurements
and the total acquisition time may be drastically reduced compared
to the present state of the art. These results will therefore allow
MRI to advance towards unravelling a wealth of diagnostic information. 
\end{abstract}
\maketitle
Information on compartment sizes and geometrical features of microstructures
in live tissues is one of the potential keys to diagnosing tissue
changes at early stages of pathologies and understanding organ malfunctioning
due to diseases. For example, the biophysical mechanisms of cancer
development and treatment are revealed by microstructure details \citep{Patterson2008,Padhani2009,White2014,onbehalfoftheMAGNIMSstudygroup2015}.
Another case where small structural changes are important indicators
are neuronal diseases that alter the distribution of axon diameters
and myelin sheath, and thereby the speed of information propagation
in the white matter of brain \citep{Hursh1939,Waxman1972,Drago2011,White2013,Xu2014,Grussu2017}.
Such diseases include Alzheimer, autism, amyotrophic lateral sclerosis
and schizophrenia. Therefore, a major goal of medical diagnosis is
the development of precise and non-invasive techniques for characterizing
the distribution of axon diameters in the brain and the sizes of microstructure
compartments in tissues \citep{Assaf2008,Alexander2010,Shemesh2015}.
In order to find reliable biomarkers based on quantitative characterization
of tissue microstructure, the diagnostic tools should provide precise
measures of tissue structure size of the order of a few micrometers.
For this purpose, it is not necessary to obtain micron-scale images
of individual tissue compartments, but it is important to measure
their\emph{ average} sizes at this resolution.

Magnetic resonance imaging (MRI) is an excellent tool for such studies,
since it enables detailed, non-invasive characterization of tissues
in vivo. Its resolution, in terms of voxel sizes, is typically limited
to millimeters in clinical studies, or hundreds of microns in preclinical
studies but reaching micrometer scales under specific conditions \citep{Lambert:2009kx,MOORE20151}.
However, it also offers the potential to quantify structural details
that are orders of magnitude smaller than the size of a voxel by monitoring
the distance over which water molecules can travel by diffusion until
their motion is restricted by walls that are not directly visible,
such as cellular membranes. This approach is often called diffusion-weighted
imaging (DWI) \citep{Stejskal1965,LeBihan2003,Grebenkov2007,Callaghan2011}.
The most promising DWI technique employs Modulated Gradient Spin Echo
(MGSE) sequences that enable detailed microstructure characterization
\citep{Stepisnik1993,Callaghan1995,Callaghan1997,Shemesh2013,Drobnjak2016,Nilsson2017,Kakkar2018}.
Several works have addressed the estimation of compartment sizes by
protocols based on various DWI sequences \citep{Ong2010,Komlosh2011,Alvarez2013a,Shemesh2013,Drobnjak2016,Nilsson2017,Kakkar2018,Xu2019}.
As pointed out by those works, the main open questions are: what resolution
can be ultimately achieved by these experiments and how the experimental
parameters should be adapted to approach this ultimate limit? 

In this work, we provide answers to these important questions by adapting
important results from quantum-information theory: we analytically
obtain this limit for the important case where the size of tissue
microstructures are probed by diffusion processes via DWI experiments.
We derive the necessary control conditions for MGSE sequences to allow
the attainment of this limit. As examples, we consider sequences with
typical modulated gradient waveforms for the estimation of microstructure
compartment sizes, assuming generic geometries. We show that the ultimate
precision limit of the estimation is achievable by MGSE sequences
if the gradient strength is set to a value that depends on the microstructure
size, the $T_{2}$ relaxation time and the diffusion coefficient of
the molecules within the compartments. Based on this result, an optimization
protocol is provided for MGSE sequences capable of attaining the highest
possible precision under the given experimental constraints. With
this protocol, the total acquisition time of quantitative microstructure
imaging is shown to be \emph{drastically reduced}, taking into account
the limitations of the available hardware and the microscopic properties
of the tissues being studied, such as diffusion constants and spin
relaxation times.
\begin{figure*}
\includegraphics[width=1\textwidth]{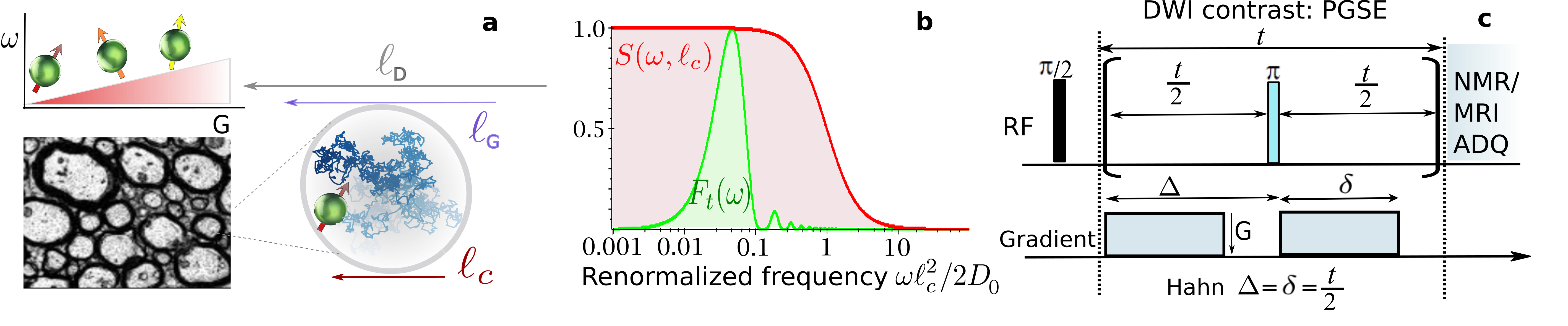}

\caption{{\label{fig-Fw-Gw}}\textbf{Diffusion-weighted spin-echo probing
of length scales in tissue microstructure.} \textbf{(a)} Schematic
image of axon diameters in brain white matter. The myelin sheaths
(dark circles) protect the cells of the central nervous system. In
the presence of a magnetic field gradient, the Larmor frequency of
the nuclear spins depends on their spatial positions. As the spins
undergo Brownian motion, their displacement is limited by the distance
between the walls of the compartment. Three relevant scales are compared:
the restriction length $\ell_{c}$ is the main quantity that we have
to characterize; it is related to the size of the compartment. The
diffusion length $\ell_{D}$ is the average distance that the water
molecules would travel in the case of free diffusion and $\ell_{G}$
is a dephasing length scale determined by experimental parameters
that are sensitive to the diffusion process, such as the amplitude
$G$ of the gradients. \textbf{(b)} Scheme of a normalized displacement
power spectrum $S(\ell_{c},\omega)$ (red color) and an optimal MGSE
filter function $F_{t}(\omega)$ (green color) as a function of the
renormalized frequency $\omega\ell_{c}^{2}/2D_{0}$. \textbf{(c)}
Pulsed Gradient Spin Echo (PGSE) sequence for probing the diffusion
restriction length: an initial $\frac{\pi}{2}$-excitation pulse is
followed by the diffusion weighting-contrast building block that consists
of a refocusing ($\pi$)-pulse at half the diffusion weighting time
$t$ between two pulsed gradients of duration $\delta$ and separated
by a delay $\text{\ensuremath{\Delta}}$. After the diffusion-weighting
period, the remaining signal is measured, possibly using MRI encoding.
The PGSE filter shown in panel \textbf{b}, assumes $\delta=\Delta=\frac{t}{2}$.\label{fig-Sequences-control-1}}
\end{figure*}

\textbf{Diffusion{-weighted spin-echo NMR signal}.} Nuclear spins
of molecules in biological tissues, in particular the spins $s=\frac{1}{2}$
of protons in water molecules, interact with external magnetic fields
in MRI. A uniform magnetic field along the $z$ axis defines the Larmor
precession frequency of the spins. In DWI, uniform magnetic-field
gradients $G\hat{r}$ are applied along an arbitrary direction $\hat{r}$
(Fig. \ref{fig-Fw-Gw}a). The spins are then subjected to fluctuating
precession frequencies $\omega(t)$ induced by molecular diffusion
displacements \citep{Grebenkov2007,Callaghan2011}. The instantaneous
precession frequency of the spin $\omega(t)=\gamma Gr(t)$ accounts
for the random motion of molecular diffusion, where $r(t)$ is the
instantaneous position of the diffusing spin along the field-gradient
direction and $\gamma$ is the gyromagnetic factor of the nucleus.
If the motion of the molecules is restricted, e.g. due to compartmentalization
of the tissue, such as in the human brain, the fluctuations are limited
in amplitude. Quantification of these fluctuations therefore allows
one to obtain an indirect measure of the size distribution of these
compartments as shown in Fig. \ref{fig-Fw-Gw}a \citep{Assaf2008,Alexander2010,Shemesh2015}.

A typical MRI / DWI experiment starts with a $\pi/2$-excitation pulse,
which creates a coherent superposition of the two spin states. During
the subsequent evolution within the diffusion time $t$, the spin-ensemble
magnetization $M(t)=\left\langle e^{-i\phi(t)}\right\rangle M(0)$
records the diffusion process, where the brackets represent the ensemble
average over the random phases. The total magnetization of the sample
is encoded spatially by an MRI sequence at the end of this evolution.
We consider here that $M(t)$ is the magnetization in a voxel of the
image, that provides what is called the diffusion-weighted contrast.
A Gaussian phase distribution is typically assumed for the random
phase $\phi(t)$, leading, under probing by a MGSE sequence, to a
decay of the magnetization \citet{Stepisnik1999}
\begin{equation}
M(t)=e^{-\frac{1}{2}\left\langle \phi^{2}(t)\right\rangle }M(0),\label{eq:M(TE)}
\end{equation}
with the mean value $\left\langle \phi(t)\right\rangle =0$. Here,
the signal attenuation factor $\frac{1}{2}\left\langle \phi^{2}(t)\right\rangle $
accounts for the contrast generated in an image, based on the effect
of the MGSE sequence that probes different time scales of the diffusion
process. The phase variance can be described in the frequency domain
by the following expression \citep{Callaghan1995,Stepisnik2006,Lasic2006,Alvarez2011a},
which is the example for the case of phase-diffusion of the Kofman-Kurizki
(KK) universal formula for decoherence (dephasing and relaxation)
control and probing in quantum systems \citep{kofman2000acceleration,kofman_universal_2001,kofman_unified_2004,gordon_universal_2007,Alvarez2011a,Kurizki2015a,Zwick2016a},
\begin{equation}
\left\langle \phi^{2}(t)\right\rangle =\gamma^{2}\int_{-\infty}^{\infty}d\omega F_{t}(\omega)S(\omega).\label{eq:quadraticdisplacement-1}
\end{equation}
The phase variance is a convolution of two spectral functions: (i)
The filter function $F_{t}(\omega)$ is the finite-time Fourier transform
(FT) of the magnetic field gradient (power spectrum) applied in the
experimental sequence. It acts as a spectral noise-filter: If its
value is 1, it passes the experimental noise at that frequency without
attenuation whereas if it is 0, it blocks the noise completely. (ii)
The spectral density $S(\omega)$ of the spin noise induced by the
environment is given (in the case of diffusion) by the FT of the spin
displacement autocorrelation function $\left\langle \Delta r(t)\Delta r(t+\tau)\right\rangle $,
where $\Delta r(t)=r(t)-\left\langle r(t)\right\rangle $ is the instantaneous
displacement deviation from the mean value \citep{Lasic2006,Stepisnik2006}.
For molecular diffusion $\left\langle \Delta r(t)\Delta r(t+\tau)\right\rangle =D_{0}\tau_{c}e^{-|\tau|/\tau_{c}}$,
where $D_{0}$ is the free diffusion coefficient \citep{Klauder1962,Alvarez2013a,Shemesh2013}
and $\tau_{c}$ is the correlation time. For molecules that diffuse
in a microstructure, the characteristic time $\tau_{c}$ is the one
required on average for a molecule to probe the compartment boundaries.
It is related to the restriction length by Einstein's expression $\ell_{c}^{2}=2D_{0}\tau_{c}$
\citep{Callaghan2011}. The spectral density $S(\omega)$ is given
by \citep{Klauder1962,Stepisnik1993,Lasic2006,kofman_unified_2004,Zwick2016a}
\begin{equation}
{\color{red}{\color{black}S(\omega)=\frac{D_{0}\tau_{c}^{2}}{\pi(1+\omega^{2}\tau_{c}^{2})}.}}\label{eq:Difussion-Spectrum_S(l_c)}
\end{equation}
Figure \ref{fig-Fw-Gw}b shows $F_{t}(\omega)=\frac{1}{2\pi}\left|\intop_{0}^{t}dt'G(t')e^{-i\omega t'}\right|^{2}$
and $S(\omega)$ for a typical MGSE sequence displayed in Fig. \ref{fig-Fw-Gw}c.
In restricted diffusion, the specific relation between $\ell_{c}$
and the geometric size depends on the compartment shape (see Methods);
e.g., for cylinders oriented perpendicular to the direction of the
magnetic field gradient, a good approximation is $\ell_{c}=0.37d$,
where $d$ is the cylinder diameter \citep{Stepisnik1993,Callaghan1995,Stepisnik2006,Shemesh2013}.

\textbf{Ultimate error bounds for estimating microstructure sizes.}
The central question we pose is: What is the best MGSE control strategy
to infer the restriction length of the diffusion process? To answer
this question we resort to quantum information tools, in order to
determine optimal gradient control strategies for obtaining the \emph{best
estimation of the restriction length of the diffusion process}, and
thereby determine microstructure sizes in biological tissues.

The figure of merit for the estimation of $\ell_{c}$, the parameter
that determines microstructure sizes, is the relative error $\frac{\delta\ell_{c}}{\ell_{c}}$.
Assuming unbiased single-parameter estimation, the relative error\emph{
}of the magnetization signal in Eq. (\ref{eq:M(TE)}) is limited by
the Cramer-Rao bound
\begin{equation}
\frac{\delta\ell_{c}}{\ell_{c}}\geq\frac{\varepsilon(t,\ell_{c})}{\mathcal{\sqrt{N}}}=\frac{1}{\ell_{c}\sqrt{\mathcal{\mathcal{N}F_{Q}}(t,\ell_{c})}},\label{eq:rel error-1-1}
\end{equation}
where \emph{$\varepsilon$ is the minimal attainable relative error
per measurement}, which is determined by the quantum Fisher information
(QFI) $\mathcal{F_{Q}}(t,\ell_{c})$ of $\ell_{c}$ obtainable from
the measured spin and $\mathcal{N}$ is the number of measurements
\citep{Paris2009_QUANTUM-ESTIMATION,Caves_1994_fisher,cramer1999mathematical}.
The QFI depends on the magnetization signal in Eq. (\ref{eq:M(TE)})
\citep{Paris2009_QUANTUM-ESTIMATION,Paris2014_Characterization-of-classical-Gaussian,Zwick_PhysScrip_2015,Zwick2016a}
(see Methods) with a functional dependence on $\ell_{c}$ and the
total diffusion weighting time $t$. This expression implicitly depends
on other system parameters, particularly the diffusion coefficient
$D_{0}$ that we assume to be known.

The relative error per measurement \emph{$\varepsilon$}, Eq. ({\ref{eq:rel error-1-1}}),
can be minimized by maximizing $\mathcal{F_{Q}}$ with respect to
the MGSE control parameters, specifically the gradient strength and
modulation shape \citep{Zwick2016a}. For a given MGSE sequence and
a given gradient strength $G$, the optimal diffusion weighting time
$t_{opt}$ is defined by
\begin{equation}
\mathcal{F_{Q}}(t_{opt},\ell_{c})=\underset{t}{\mathrm{max}}(\mathcal{F_{Q}}(t,\ell_{c})).\label{eq:t_opt}
\end{equation}
This define a minimal error $\varepsilon(t_{opt},\ell_{c})$ for each
MGSE control.

We can see that by suitably designing an optimal MGSE control, one
can \emph{attain} the \emph{ultimate relative-error bound for the
restriction length of the diffusion process} (see Methods), namely,
\begin{equation}
\varepsilon(t,\ell_{c})\ge\varepsilon_{0}.\label{eq:ultimate error bound}
\end{equation}
 Remarkably, this precision estimation bound for $\ell_{c}$ is general
for all possible MGSE control sequences and \emph{independent of the
particular geometry} restricting the diffusion (see Methods). This
ultimate precision is attained by optimally choosing the following
length scales (Fig. \ref{fig-Fw-Gw}a):
\begin{equation}
\ell_{G}=\sqrt[3]{\frac{2D_{0}}{\gamma G}}\quad\ell_{c}=\sqrt{2D_{0}\tau_{c}}\quad\ell_{D}=\sqrt{2D_{0}t},\label{eq:grad-corr-diff-lenght}
\end{equation}
where $\ell_{G}$ the dephasing length, $\ell_{c}$ the restriction
length, and $\ell_{D}$ is the the diffusion length.

\textbf{Attaining the ultimate precision bound.} To attain the ultimate
error bound per measurement in Eq. (\ref{eq:ultimate error bound}),
the MGSE should satisfy the following requirements: \emph{(i)} The
spectral filter $F_{t}(\omega)$ should overlap with the displacement
power spectrum $S(\ell_{c},\omega)$ (Eq. (\ref{eq:Difussion-Spectrum_S(l_c)}))
within the spectral region of the highest power-law dependence on
$\ell_{c}$ at a \emph{low frequency}, where \emph{$S(\ell_{c},\omega\approx0)\propto\ell_{c}^{4}$,
}corresponding to $-\ln\left[M(t)/M(0)\right]\propto\ell_{c}^{4}$.\emph{
(ii)} The total diffusion weighting time $t$ should be the optimal
time $t_{opt}$, Eq. (\ref{eq:t_opt}), such that $\ln\left[M(t_{opt},\ell_{c})/M(0)\right]=\ln M_{o}$,
where the optimal magnetization contrast $M_{o}$ is determined by
the expression $-\ln M_{o}=1+W(-2e^{-2})\approx0.8$ (see Methods).
A MGSE sequence producing a narrow \emph{low-frequency} bandpass filter
would be therefore the most sensitive to the restriction size $\ell_{c}$.
The narrowest \emph{low-frequency} bandpass filter for a given diffusion
time among typical MGSE sequences is an optimized Pulse Gradient Spin
Echo (PGSE) sequence, as it only contains one gradient-sign switch
leading to the longest possible modulation period. The PGSE sequence
\citep{Stejskal1965} is displayed in Figure \ref{fig-Sequences-control-1}c.
The time $\delta$ is the gradient pulse duration and $\Delta$ is
the delay between the gradient pulses. The PGSE sequence includes
one $\pi$ rf-pulse to refocus external magnetic field inhomogeneities.
PGSE is analogous to a gradient echo or a Hahn spin echo \citep{Hahn1950}
with a constant gradient if $\delta=\Delta=\frac{t}{2}$, where $t$
is the sequence duration and therefore the diffusion time (Fig. \ref{fig-Sequences-control-1}c)
and produces the lowest frequency bandpass under this limit.

If $t\gg\tau_{c}$, the condition \emph{(i)} to attain the bound is
satisfied resulting in (see Methods)
\begin{eqnarray}
-\ln\left(\frac{M_{\delta=\Delta=\frac{t}{2},t\gg\tau_{c}}(t,\ell_{c})}{M(0)}\right) & \approx & \gamma^{2}G^{2}D_{0}\tau_{c}^{2}t\propto\ell_{c}^{4}.\label{eq:Hahn-decay}
\end{eqnarray}
From requirement (ii) and Eqs. (\ref{eq:Hahn-decay}), the optimal
diffusion time is found to satisfy
\begin{equation}
t_{opt}=\frac{-\ln M_{o}}{\gamma^{2}G^{2}D_{0}\tau_{c}^{2}}\approx\frac{0.8}{\gamma^{2}G^{2}D_{0}\tau_{c}^{2}}.\label{eq:topt hahn}
\end{equation}
\begin{figure}
\includegraphics[width=1\columnwidth]{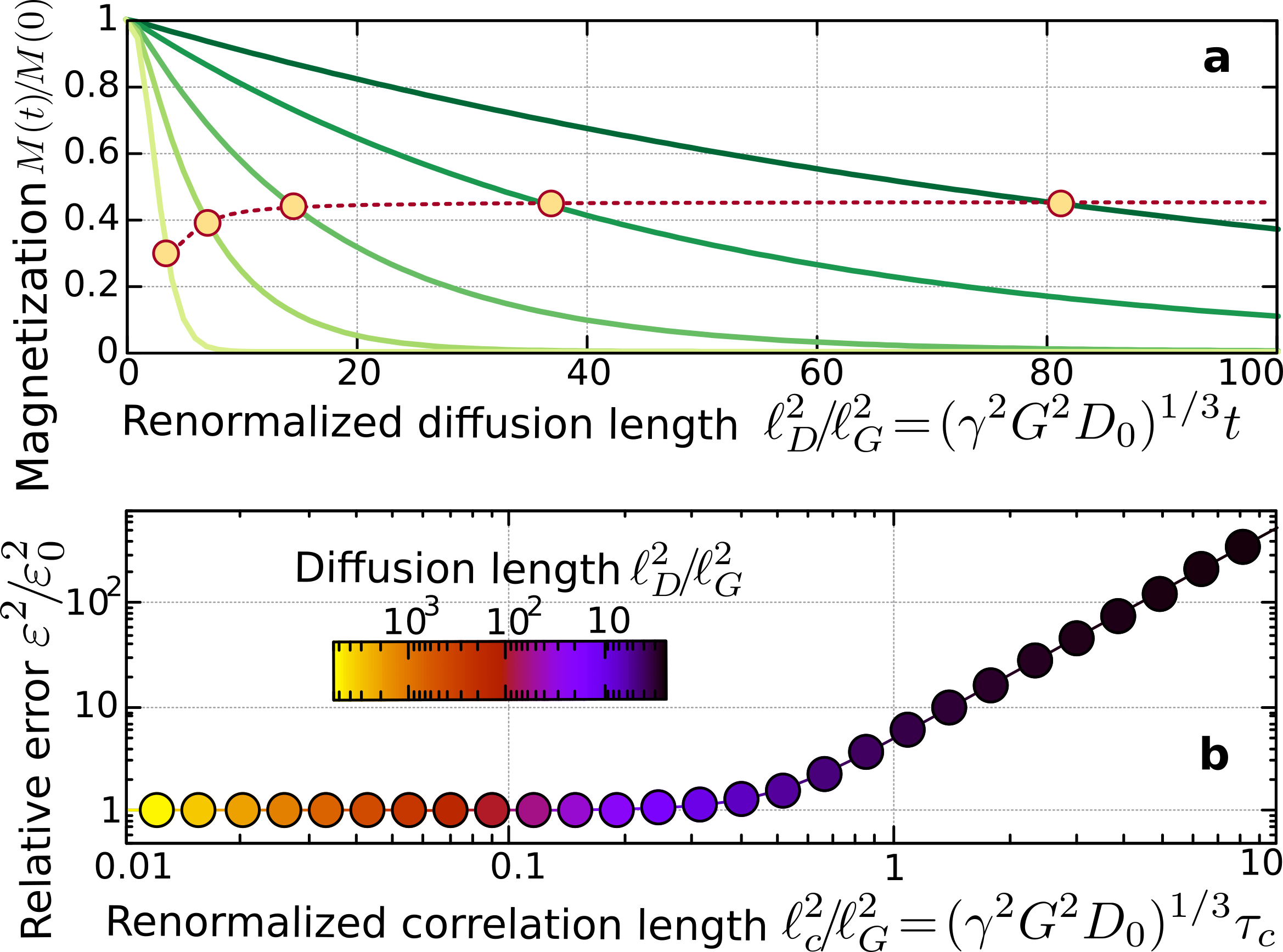}

\caption{{\label{fig-Magnetization}}\textbf{Attaining the ultimate precision
bound of the restriction length }$\ell_{c}$ \textbf{from diffusion-weighted
magnetization decay. (a)} The decay of the normalized signal corresponding
to the Hahn MGSE control $M(t)/M(0)$ as a function of the diffusion
time in units of the square of the normalized diffusion length $\ell_{D}^{2}/\ell_{G}^{2}$,
with $\ell_{D}^{2}=2D_{0}t$. Dark to light green lines denote increasing
renormalized correlation times in units of the renormalized restriction
lengths $\ell_{c}^{2}/\ell_{G}^{2}\!=\!\left(\gamma^{2}G^{2}D_{0}\right)^{1/3}\tau_{c}\!=0.1,0.15,0.25,0.4,1$.
The circles denote the optimal diffusion times $\left(\gamma^{2}G^{2}D_{0}\right)^{1/3}t_{opt}$.
\textbf{(b)} Relative minimum error $\frac{\varepsilon^{2}}{\varepsilon_{0}^{2}}$
in the estimation of $\ell_{c}$ per measurement, i.e. $\mathcal{N}=1$,
as a function of the renormalized restriction length $\ell_{c}^{2}/\ell_{G}^{2}$
(colored circles). The ultimate error bound in the estimation of $\ell_{c}$,
$\varepsilon_{0}$ in Eq. (\ref{eq:ultimate error bound}), is set
at unity. The colored scale for the circles gives the normalized diffusion
length $\ell_{D}^{2}/\ell_{G}^{2}$. The bound is attained when $\ell_{c}^{2}/\ell_{G}^{2}\ll1$
and $\ell_{D}^{2}/\ell_{G}^{2}\gg1,$ and $\frac{\ell_{c}^{4}}{\ell_{G}^{4}}\frac{\ell_{D}^{2}}{\ell_{G}^{2}}=-\ln M_{0}$
as described in Eq. (\ref{eq:topt hahn}), implying $\left(\gamma^{2}G^{2}D_{0}\right)^{1/3}\tau_{c}\ll1$.
The quantity $\frac{\varepsilon^{2}}{\varepsilon_{0}^{2}}$ yields
directly the \emph{number of measurements }$\mathcal{N}$ \emph{needed
to attain the ultimate precision bound per measurement}.}
\end{figure}
Figure \ref{fig-Magnetization}a shows typical signal decay for different
$\gamma^{2}G^{2}D_{0}\tau_{c}^{3}$ values and the corresponding optimal
diffusion times $t_{opt}$. Both requirements \emph{(i)} and \emph{(ii)}
are therefore fulfilled when $t=t_{opt}\gg\tau_{c}$, meaning that
the diffusion length should be much larger than the correlation length,
$\ell_{D}\gg\ell_{c}$ (Eq. \ref{eq:grad-corr-diff-lenght}). Together
with Eq. (\ref{eq:topt hahn}), this requirement amounts to the condition
$\gamma^{2}G^{2}D_{0}\tau_{c}^{3}\ll1$ which means that the diffusion
correlation length should be much smaller than the dephasing length,
$\ell_{c}\ll\ell_{G}$ (see Fig. \ref{fig-Fw-Gw}a). Indeed, the optimal
diffusion-weighted length $\ell_{D}/\ell_{G}$ that corresponds to
the highest precision per measurement for determining the restriction
length, is seen from Fig. \ref{fig-Magnetization}b to require
\begin{equation}
\frac{\ell_{c}^{6}}{\ell_{G}^{6}}=\gamma^{2}G^{2}D_{0}\tau_{c}^{3}\ll1.\label{eq:high-precission-condition}
\end{equation}
Therefore, under the idealized relaxation-free condition discussed
here, \emph{increasing} \emph{$\ell_{G}$}, i.e.\emph{ reducing the
gradient strength, always improves the precision} up to the point
allowing to attain the bound\emph{ }(Eq. \ref{eq:ultimate error bound}).
Since the optimal time must fulfill Eq. (\ref{eq:topt hahn}), it
must satisfy the power-law dependence $\frac{t_{opt}}{\tau_{c}}=\frac{\ell_{D}^{2}}{\ell_{G}^{2}}\frac{\ell_{G}^{2}}{\ell_{c}^{2}}\approx\left(\gamma^{2}G^{2}D_{0}\tau_{c}^{3}\right)^{-1}=\frac{\ell_{G}^{6}}{\ell_{c}^{6}}\gg1$.
By contrast, when $\frac{t_{opt}}{\tau_{c}}\lesssim1$, $t_{opt}$
saturates at a diffusion time value where the signal decays below
$M/M(0)\approx1/e$ (Fig. \ref{fig-Magnetization}a). The larger $\tau_{c}$,
the shorter is $t_{opt}$ in Eq. (\ref{eq:topt hahn}), and the restricted
diffusion regime $t\gg\tau_{c}$ can no longer be achieved.

Figure \ref{fig-Magnetization}b shows the minimal relative squared
error $\varepsilon^{2}$ scaled to $\varepsilon_{0}^{2}$ per measurement,
$\frac{\varepsilon^{2}}{\varepsilon_{0}^{2}}$. This scaled squared
error determines the \emph{number of measurements }$\mathcal{N}$
\emph{needed to attain an error equivalent to the ultimate precision
per measurement} (see Eq. (\ref{eq:rel error-1-1})). For $\frac{\ell_{c}}{\ell_{G}}>1$,
the restricted diffusion regime is no longer achieved and the relative
error linearly increases with $\frac{\ell_{c}}{\ell_{G}}$.

\textbf{Precision bounds with transverse relaxation.} Under the idealized
relaxation-free conditions discussed so far, by reducing the gradient
we may always increase the optimal diffusion time so as to achieve
the ultimate precision bound for $\ell_{c}$ estimation. However,
this approach may fail, as the intrinsic nuclear-spin $T_{2}$-relaxation
limits the accessible diffusion probing time. The $T_{2}$-relaxation
contributes a global attenuation factor to the signal decay, which
is independent of the MGSE sequence and the corresponding diffusion
weighting. The echo signal of Eq. (\ref{eq:M(TE)}) is then $M_{T_{2}}(t,\ell_{c})=e^{-\frac{t}{T_{2}}}M(t,\ell_{c})$,
where $M(t,\ell_{c})$ accounts for the diffusion weighted spins'
magnetization of Eq. (\ref{eq:M(TE)}). We can see that the relative
error including the $T_{2}$-relaxation effects is bounded by 
\begin{equation}
\varepsilon\ge e^{\frac{t}{T_{2}}}\varepsilon_{0}\label{eq:ultimate  error bound-T2}
\end{equation}
 and thus exponentially increases with $\frac{t}{T_{2}}$ (See Methods).

The conditions for attaining the ultimate error bound are now more
restrictive, since the diffusion time $t$ cannot be larger than $T_{2}$.
This condition implies specific values for the optimal diffusion time
$t_{opt}$ and the efficiency parameter $\gamma^{2}G^{2}D_{0}\tau_{c}^{3}$
that can attain the best precision limit in the estimation of $\ell_{c}$.
This is in contrast to the limit $T_{2}\rightarrow\infty$, where
a semi-infinite range of $t_{opt}$ values exists for $\gamma^{2}G^{2}D_{0}\tau_{c}^{3}\ll1$,
that allow the ultimate error bound to be attained (see Fig. \ref{fig-Magnetization}).

\begin{figure}[!h]
\includegraphics[width=1\columnwidth]{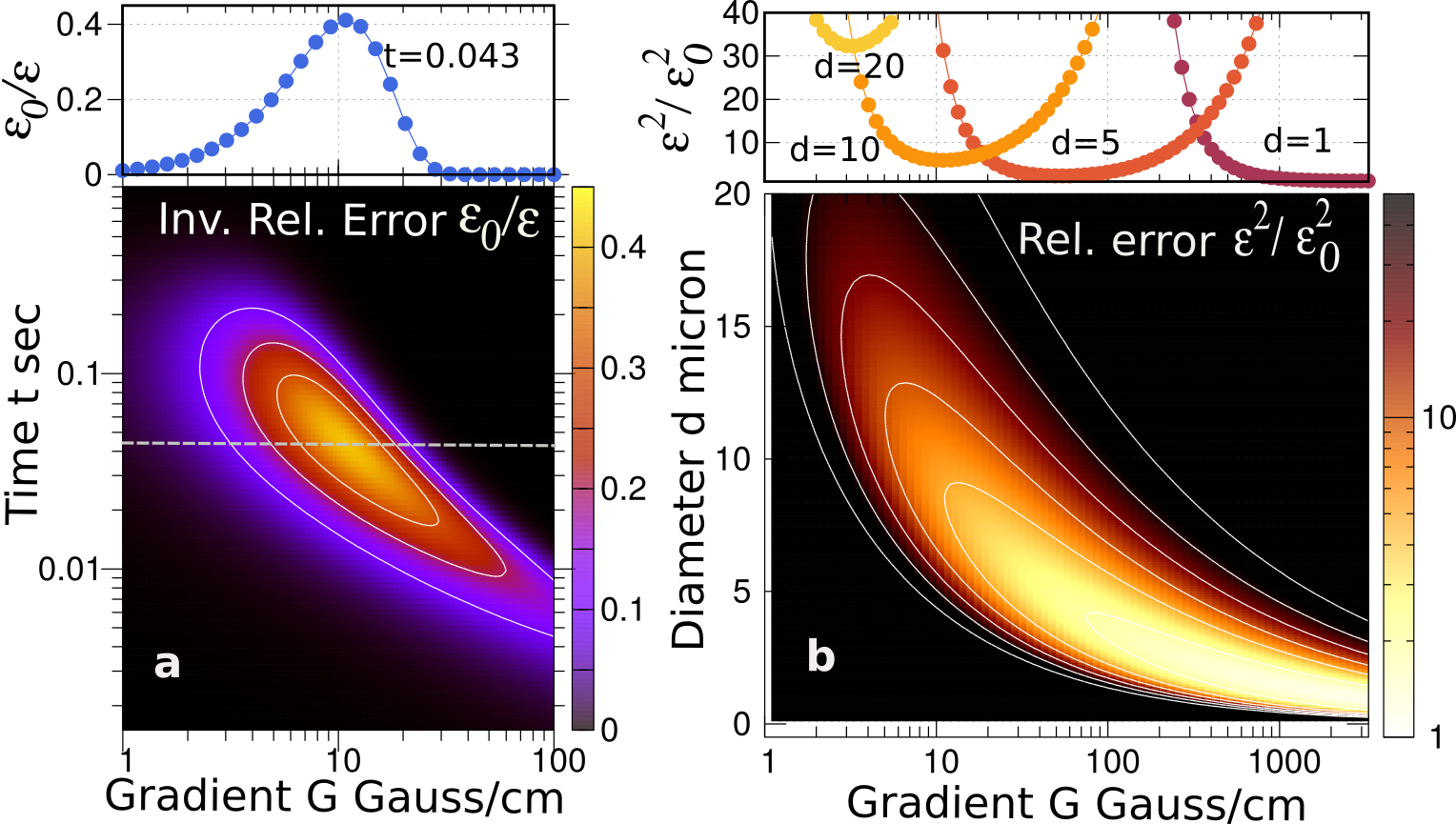}

\caption{{\label{fig-examples-diameters_d1-5-10-20_water-WGmatter}\textbf{Attaining
the precision bounds of the restriction length }$\ell_{c}$\textbf{
with transverse relaxation.} \textbf{(a)} Effects of $T_{2}$-relaxation
on the inverse of the estimation error $\frac{\varepsilon_{0}}{\varepsilon}$
in the estimation of $\ell_{c}$ per measurement, i.e. $\mathcal{N}=1$,
highlighting the optimal time and gradient strength for a cylindrical
compartment diameter of $d=10\,\mathrm{\mu m}$. This geometrical
shape is representative of axons or microfibers in general. The diffusion
coefficient $D_{0}=1\times10^{-5}\,\mathrm{\frac{cm^{2}}{s}}$ and
the relaxation time $T_{2}=0.1\:\mathrm{s}$ are typical of gray/white
matter in the brain. The correlation length is $\ell_{c}=0.37d$.
The top inset shows the inverse of the estimation error $\frac{\varepsilon_{0}}{\varepsilon}$
as a function of the gradient $G$ for a diffusion time $t=43$ms,
shown by the white dashed line in the main panel. \textbf{(b)} The
optimal squared relative error in the estimation of $d$ is normalized
to the squared minimal error bound per measurement, $\varepsilon^{2}/\varepsilon_{0}^{2}$,
is plotted as a function of the gradient $G$ and diameter $d$. The
optimal gradient strengths are shown by brighter colors for a given
diameter $d$. The relaxation time $T_{2}$ and the diffusion coefficient
$D_{0}$ are considered as in \textbf{a}. The top inset shows the
$\varepsilon^{2}/\varepsilon_{0}^{2}$ for $d=1,5,\,10,\,20\,\mathrm{\mu m}$.
The number of measurements required to attain $\varepsilon_{0}$ under
the optimal conditions grows as the diameter $d$ increases or $T_{2}$
decreases. Equivalently, the range of optimal $G$ values is reduced
as $\tau_{c}=\tfrac{1}{2}\ell_{c}^{2}/D_{0}$ approaches $T_{2}$
(Eq. \ref{eq:eq:G_conditions_under_T2}).}}
\end{figure}

Yet, upon shortening $T_{2}$, while keeping $t_{opt}\ll T_{2}$ ($\ell_{D}\ll\ell_{T_{2}}=\sqrt{2D_{0}T_{2}}$
as Eq. (\ref{eq:grad-corr-diff-lenght})), we still find a \emph{finite
region} where $\varepsilon(t_{opt},\ell_{c})\approx\varepsilon_{0}$.
This region is defined by the condition $\frac{\ell_{c}^{6}}{\ell_{G}^{6}}\ll1$,
Eq. (\ref{eq:high-precission-condition}), which has been imposed
to satisfy the requirements for achieving the ultimate error bound.
These two conditions imply that the ultimate error bound can be approached
only if $\frac{\ell_{G}}{\ell_{T_{2}}}\ll1$. Therefore the range
of restriction lengths $\ell_{c}$ that can be determined efficiently
with DWI must obey 
\begin{equation}
\left(\ell_{G}/\ell_{T_{2}}\right)\ell_{G}^{2}\ll\ell_{c}^{2}\ll\ell_{G}^{2},
\end{equation}
which are limited by the achievable gradient strengths and the $T_{2}$-relaxation
time. For a given $\ell_{c}$, these conditions define the optimal
gradients for estimating $\ell_{c}$
\begin{equation}
\frac{1}{\gamma\ell_{c}^{2}}\sqrt{\frac{2D_{0}}{T_{2}}}\ll G\ll\frac{2D_{0}}{\gamma\ell_{c}^{3}}.\label{eq:eq:G_conditions_under_T2}
\end{equation}
The minimum relative error for different values of $G$ and $\ell_{c}$
is shown in Fig. \ref{fig-examples-diameters_d1-5-10-20_water-WGmatter},
highlighting the optimal values of $G$ for estimating the allowed
range of restriction lengths.

\textbf{Conclusions.} Precise measurements of the size of microscopic
tissue compartment, such as the diameter of neuronal axons is an ongoing
endeavor that promises improved diagnostic value for many medical
conditions. Diffusion of water molecules in these tissues provides
a built-in tool that can be accessed by diffusion-weighted magnetic
resonance techniques. The present analysis, which uses tools developed
in quantum information science, reveals a universal, ultimate precision
limit for estimating microstructure sizes by DWI. This limit is attainable
by current MRI techniques available in many clinical settings, provided
the relevant control parameters are properly chosen. We have shown
here how the optimal parameters depend on the diffusion coefficient
of the probe molecules (typically water), the relaxation time $T_{2}$,
and the expected restriction length. The results of the present analysis
are very encouraging as they show that optimal estimation of microstructure
detail, e.g. axon diameters ($\sim0.1-20\,\mu m$) is achievable by
present technologies, given that modern clinical magnets can apply
gradients of hundreds Gauss/cm \citep{Setsompop2013}, and preclinical
micro-imaging magnet can apply thousands of Gauss/cm. Even smaller
microstructure sizes can, in principle, be determined with high precision,
but this may demand higher field gradients that are currently unavailable
in clinical magnets. The present results advance towards designing
quantitative and precision imaging approaches opening new avenues
for characterizing tissue microstructures in the shortest time possible,
which is imperative to find useful biomarkers for medical diagnosis.

\subsection*{Acknowledgment}

We thank L. Frydman and J. Jovicich for fruitful discussions. This
work was supported by the EU FET Open PATHOS (G.K. and D.S.); DFG
FOR 7024, ISF, QUANTERA PACE-IN (G.K.); CONICET, Instituto Balseiro
and CNEA (A.Z., G.A.A.); ANPCyT-FONCyT PICT-2017-3447, PICT-2017-3699,
PICT-2018-04333, PIP-CONICET (11220170100486CO), UNCUYO SIIP Tipo
I 2019-C028 (A.Z., G.A.A.).

\medskip{}

\section*{Methods}

\textbf{Quantum Fisher Information on the microstructure size }$\ell_{c}$\textbf{.}
The quantum Fisher information (QFI) is given by \citep{Paris2009_QUANTUM-ESTIMATION,Paris2014_Characterization-of-classical-Gaussian,Zwick_PhysScrip_2015,Zwick2016a}
\begin{eqnarray}
\mathcal{F_{Q}}(t,\ell_{c}) & = & \frac{\left[\frac{M(t,\ell_{c})}{M(0)}\right]^{2}}{1-\left[\frac{M(t,\ell_{c})}{M(0)}\right]^{2}}\left(\frac{\partial\ln\left[\frac{M(t,\ell_{c})}{M(0)}\right]}{\partial\ell_{c}}\right)^{2},\label{eq:QFI_lc}
\end{eqnarray}
with explicit functional dependence on $\ell_{c}$ and the total diffusion
weighting time $t$, but implicitly depends on the diffusion coefficient
$D_{0}$ that we assume to be known, determined by monitoring the
free diffusion time scale.

\textbf{Ultimate error bound on $\ell_{c}$. }The QFI is maximized
at $t_{opt}$ which provides the best tradeoff between the amplitude
contrast of the diffusion-weighted echo-signal ($M^{2}\left(M^{2}(0)-M^{2}\right)^{-1}$)
and its parametric-sensitivity to $\ell_{c}$, $\left|\frac{\partial\ln\left(M/M(0)\right)}{\partial\ell_{c}}\right|^{2}$.
This parametric-sensitivity depends on $\frac{\partial S}{\partial\ell_{c}}$
since the correlation length $\ell_{c}$ only appears in the displacement
power spectrum, Eq. (\ref{eq:Difussion-Spectrum_S(l_c)}). It is bounded
by $\left|\frac{\partial S}{\partial\ell_{c}}\right|\le\frac{4S}{\ell_{c}}$.
This bound is only reached when low frequencies are probed by the
MGSE filter function. The spectrum in Eq. (\ref{eq:Difussion-Spectrum_S(l_c)})
then becomes an homogeneous function of $\ell_{c}$ of order $4$,
i.e. $S(\ell_{c},\omega\approx0)\propto\ell_{c}^{4}.$ This dependence
leads to the bound
\begin{equation}
\left|\frac{\partial\ln\left(M/M(0)\right)}{\partial\ell_{c}}\right|\le\frac{4(-\ln\left(M/M(0)\right))}{\ell_{c}},\label{eq:bound-derlnM}
\end{equation}
which is attained when the MGSE control is such that it generates
a narrow \emph{low-frequency} bandpass filter. This analysis determines
a tight lower bound for the relative error in Eq. (\ref{eq:rel error-1-1})

\begin{equation}
\varepsilon(t,\ell_{c})\ge\frac{\sqrt{\left[M(0)\right]^{2}-\left[M(t,\ell_{c})\right]^{2}}}{4(-\ln\left[M(t,\ell_{c})/M(0)\right])M(t,\ell_{c})},\label{eq:error bound}
\end{equation}
which is minimized when the condition
\begin{equation}
-\ln\left[\frac{M(t,\ell_{c})}{M(0)}\right]=-\ln M_{o}=1+\frac{W(-2e^{-2})}{2}\approx0.8\label{eq:M_0}
\end{equation}
is fulfilled, with $W(z)$ being the Lambert function. This demonstrate
the existence of an \emph{attainable ultimate-relative-error bound
for the restriction length of the diffusion process}

\begin{equation}
\varepsilon(t,\ell_{c})\ge\frac{\sqrt{1-M_{o}^{2}}}{4(-\ln M_{o})M_{o}}=\varepsilon_{0}\approx0.62,\label{eq:ultimate error bound-1}
\end{equation}
where $\varepsilon_{0}=\frac{\left(\text{\ensuremath{-\frac{1}{2}}}W(\text{\textminus}2e^{\text{\textminus}2})(1+\frac{1}{2}W(\text{\textminus}2e^{\text{\textminus}2})\right)^{-\frac{1}{2}}}{4}\text{\ensuremath{\approx}}\frac{2.48}{4}\text{\ensuremath{\approx}}0.62$.

The $T_{2}$ attenuation factor $e^{-\frac{t}{T_{2}}}$ of the signal
decay, introduces a crucial constraint for attaining the optimal diffusion
time that leads to the ultimate precision bound of Eq. (\ref{eq:ultimate error bound})
in the estimation of $\ell_{c}$. The relative error is now bounded
by
\begin{multline}
\varepsilon\ge\frac{\sqrt{M^{2}(0)-e^{-\frac{2t}{T_{2}}}M^{2}}}{4(-\ln\left[M/M(0)\right])e^{-\frac{t}{T_{2}}}M}\\
\geq e^{\frac{t}{T_{2}}}\frac{\sqrt{M^{2}(0)-M^{2}}}{4(-\ln\left[M/M(0)\right])M},\label{eq:ultimate  error bound-T2-1}
\end{multline}
which leads to Eq. (\ref{eq:ultimate  error bound-T2}).

\textbf{Optimized PGSE: Hahn Spin-Echo Filter.} The narrowest \emph{low-frequency}
bandpass filter from typical MGSE sequences is an optimized version
of the PGSE sequence, as it only contains one gradient sign switch,
producing the lowest frequency bandpass. The frequency filter of PGSE
is

\begin{equation}
F_{\delta,\Delta}^{PGSE}(\omega)=\left|\frac{4ie^{-\frac{\mathrm{i}\omega(\delta+\Delta)}{2}}\sin\left(\frac{\omega\delta}{2}\right)\sin\left(\frac{\omega\Delta}{2}\right)}{\omega}\right|^{2}.
\end{equation}

For $\delta=\Delta=\frac{t}{2}$, which correspond to the well known
gradient- or Hahn spin-echo refocusing sequences over the total diffusion
time $t$ (Fig. \ref{fig-Fw-Gw}), the magnetization signal is

\begin{equation}
M_{\delta=\frac{t}{2},\Delta=\frac{t}{2}}=e^{-\gamma^{2}G^{2}D\tau_{c}^{2}t\left[1-\frac{t}{\tau_{c}}\left(3+e^{-\frac{t}{\tau_{c}}}-4e^{-\frac{t}{2\tau_{c}}}\right)\right]}M(0).\label{eq:Jhahn_Lorentz}
\end{equation}
Under this conditions, the sequence produces a low-frequency narrow
bandpass filter if $t\gg\tau_{c}$, resulting in Eq. (\ref{eq:Hahn-decay}).

\textbf{Estimating restriction lengths in general geometries. }Remarkably,
the precision estimation bound for $\ell_{c}$ is general for all
possible MGSE control sequences and \emph{independent of the particular
geometry} restricting the diffusion. The spectral density is the Fourier
Transform of the diffusion correlation function derived from the solution
to the Einstein-Fick diffusion equation for a corresponding restricting
geometry \citep{Stepisnik1993,Stepisnik1999}. The spectral density
is therefore given by 
\[
{\color{black}S(\omega)=\sum_{k}\frac{D_{0}b_{k}\tau_{k}^{2}}{\pi(1+\omega^{2}\tau_{k}^{2})},}
\]
where the coefficients $b_{k}$ and correlation times $\tau_{k}$
depend on the geometry of the compartments and can be found in Ref.
\citep{Stepisnik1993} for cylinders, spheres and planar layers. When
the MGSE filter overlaps with the displacement power spectrum $S(\omega)$
within the spectral region of \emph{low frequency $S(\ell_{c},\omega\approx0)=\sum_{k}\frac{D_{0}b_{k}\tau_{k}^{2}}{\pi}\propto\ell_{c}^{4}$.
}Here the restriction length is determined by the root mean squared
correlation time $\ell_{c}^{2}=2D_{0}\sqrt{\sum_{k}b_{k}\tau_{k}^{2}}$.

\bibliographystyle{apsrev4-1}
\bibliography{bibliography}

\begin{thebibliography}{48}%
\makeatletter
\providecommand \@ifxundefined [1]{%
 \@ifx{#1\undefined}
}%
\providecommand \@ifnum [1]{%
 \ifnum #1\expandafter \@firstoftwo
 \else \expandafter \@secondoftwo
 \fi
}%
\providecommand \@ifx [1]{%
 \ifx #1\expandafter \@firstoftwo
 \else \expandafter \@secondoftwo
 \fi
}%
\providecommand \natexlab [1]{#1}%
\providecommand \enquote  [1]{``#1''}%
\providecommand \bibnamefont  [1]{#1}%
\providecommand \bibfnamefont [1]{#1}%
\providecommand \citenamefont [1]{#1}%
\providecommand \href@noop [0]{\@secondoftwo}%
\providecommand \href [0]{\begingroup \@sanitize@url \@href}%
\providecommand \@href[1]{\@@startlink{#1}\@@href}%
\providecommand \@@href[1]{\endgroup#1\@@endlink}%
\providecommand \@sanitize@url [0]{\catcode `\\12\catcode `\$12\catcode
  `\&12\catcode `\#12\catcode `\^12\catcode `\_12\catcode `\%12\relax}%
\providecommand \@@startlink[1]{}%
\providecommand \@@endlink[0]{}%
\providecommand \url  [0]{\begingroup\@sanitize@url \@url }%
\providecommand \@url [1]{\endgroup\@href {#1}{\urlprefix }}%
\providecommand \urlprefix  [0]{URL }%
\providecommand \Eprint [0]{\href }%
\providecommand \doibase [0]{http://dx.doi.org/}%
\providecommand \selectlanguage [0]{\@gobble}%
\providecommand \bibinfo  [0]{\@secondoftwo}%
\providecommand \bibfield  [0]{\@secondoftwo}%
\providecommand \translation [1]{[#1]}%
\providecommand \BibitemOpen [0]{}%
\providecommand \bibitemStop [0]{}%
\providecommand \bibitemNoStop [0]{.\EOS\space}%
\providecommand \EOS [0]{\spacefactor3000\relax}%
\providecommand \BibitemShut  [1]{\csname bibitem#1\endcsname}%
\let\auto@bib@innerbib\@empty
\bibitem [{\citenamefont {Patterson}\ \emph {et~al.}(2008)\citenamefont
  {Patterson}, \citenamefont {Padhani},\ and\ \citenamefont
  {Collins}}]{Patterson2008}%
  \BibitemOpen
  \bibfield  {author} {\bibinfo {author} {\bibfnamefont {D.~M.}\ \bibnamefont
  {Patterson}}, \bibinfo {author} {\bibfnamefont {A.~R.}\ \bibnamefont
  {Padhani}}, \ and\ \bibinfo {author} {\bibfnamefont {D.~J.}\ \bibnamefont
  {Collins}},\ }\href {\doibase 10.1038/ncponc1073} {\bibfield  {journal}
  {\bibinfo  {journal} {Nat. Rev. Clin. Oncol.}\ }\textbf {\bibinfo {volume}
  {5}},\ \bibinfo {pages} {220} (\bibinfo {year} {2008})}\BibitemShut {NoStop}%
\bibitem [{\citenamefont {Padhani}\ \emph {et~al.}(2009)\citenamefont
  {Padhani}, \citenamefont {Liu}, \citenamefont {Koh}, \citenamefont
  {Chenevert}, \citenamefont {Thoeny}, \citenamefont {Takahara}, \citenamefont
  {Dzik-Jurasz}, \citenamefont {Ross}, \citenamefont {Van~Cauteren},
  \citenamefont {Collins}, \citenamefont {Hammoud}, \citenamefont {Rustin},
  \citenamefont {Taouli},\ and\ \citenamefont {Choyke}}]{Padhani2009}%
  \BibitemOpen
  \bibfield  {author} {\bibinfo {author} {\bibfnamefont {A.~R.}\ \bibnamefont
  {Padhani}}, \bibinfo {author} {\bibfnamefont {G.}~\bibnamefont {Liu}},
  \bibinfo {author} {\bibfnamefont {D.~M.}\ \bibnamefont {Koh}}, \bibinfo
  {author} {\bibfnamefont {T.~L.}\ \bibnamefont {Chenevert}}, \bibinfo {author}
  {\bibfnamefont {H.~C.}\ \bibnamefont {Thoeny}}, \bibinfo {author}
  {\bibfnamefont {T.}~\bibnamefont {Takahara}}, \bibinfo {author}
  {\bibfnamefont {A.}~\bibnamefont {Dzik-Jurasz}}, \bibinfo {author}
  {\bibfnamefont {B.~D.}\ \bibnamefont {Ross}}, \bibinfo {author}
  {\bibfnamefont {M.}~\bibnamefont {Van~Cauteren}}, \bibinfo {author}
  {\bibfnamefont {D.}~\bibnamefont {Collins}}, \bibinfo {author} {\bibfnamefont
  {D.~A.}\ \bibnamefont {Hammoud}}, \bibinfo {author} {\bibfnamefont
  {G.~J.~S.}\ \bibnamefont {Rustin}}, \bibinfo {author} {\bibfnamefont
  {B.}~\bibnamefont {Taouli}}, \ and\ \bibinfo {author} {\bibfnamefont {P.~L.}\
  \bibnamefont {Choyke}},\ }\href {\doibase 10.1593/neo.81328} {\bibfield
  {journal} {\bibinfo  {journal} {Neoplasia}\ }\textbf {\bibinfo {volume}
  {11}},\ \bibinfo {pages} {102} (\bibinfo {year} {2009})}\BibitemShut
  {NoStop}%
\bibitem [{\citenamefont {White}\ \emph {et~al.}(2014)\citenamefont {White},
  \citenamefont {McDonald}, \citenamefont {Farid}, \citenamefont {Kuperman},
  \citenamefont {Karow}, \citenamefont {Schenker-Ahmed}, \citenamefont
  {Bartsch}, \citenamefont {Rakow-Penner}, \citenamefont {Holland},
  \citenamefont {Shabaik}, \citenamefont {Bj{\o}rnerud}, \citenamefont {Hope},
  \citenamefont {Hattangadi-Gluth}, \citenamefont {Liss}, \citenamefont
  {Parsons}, \citenamefont {Chen}, \citenamefont {Raman}, \citenamefont
  {Margolis}, \citenamefont {Reiter}, \citenamefont {Marks}, \citenamefont
  {Kesari}, \citenamefont {Mundt}, \citenamefont {Kaine}, \citenamefont
  {Carter}, \citenamefont {Bradley},\ and\ \citenamefont {Dale}}]{White2014}%
  \BibitemOpen
  \bibfield  {author} {\bibinfo {author} {\bibfnamefont {N.~S.}\ \bibnamefont
  {White}}, \bibinfo {author} {\bibfnamefont {C.~R.}\ \bibnamefont {McDonald}},
  \bibinfo {author} {\bibfnamefont {N.}~\bibnamefont {Farid}}, \bibinfo
  {author} {\bibfnamefont {J.}~\bibnamefont {Kuperman}}, \bibinfo {author}
  {\bibfnamefont {D.}~\bibnamefont {Karow}}, \bibinfo {author} {\bibfnamefont
  {N.~M.}\ \bibnamefont {Schenker-Ahmed}}, \bibinfo {author} {\bibfnamefont
  {H.}~\bibnamefont {Bartsch}}, \bibinfo {author} {\bibfnamefont
  {R.}~\bibnamefont {Rakow-Penner}}, \bibinfo {author} {\bibfnamefont
  {D.}~\bibnamefont {Holland}}, \bibinfo {author} {\bibfnamefont
  {A.}~\bibnamefont {Shabaik}}, \bibinfo {author} {\bibfnamefont
  {A.}~\bibnamefont {Bj{\o}rnerud}}, \bibinfo {author} {\bibfnamefont
  {T.}~\bibnamefont {Hope}}, \bibinfo {author} {\bibfnamefont {J.}~\bibnamefont
  {Hattangadi-Gluth}}, \bibinfo {author} {\bibfnamefont {M.}~\bibnamefont
  {Liss}}, \bibinfo {author} {\bibfnamefont {J.~K.}\ \bibnamefont {Parsons}},
  \bibinfo {author} {\bibfnamefont {C.~C.}\ \bibnamefont {Chen}}, \bibinfo
  {author} {\bibfnamefont {S.}~\bibnamefont {Raman}}, \bibinfo {author}
  {\bibfnamefont {D.}~\bibnamefont {Margolis}}, \bibinfo {author}
  {\bibfnamefont {R.~E.}\ \bibnamefont {Reiter}}, \bibinfo {author}
  {\bibfnamefont {L.}~\bibnamefont {Marks}}, \bibinfo {author} {\bibfnamefont
  {S.}~\bibnamefont {Kesari}}, \bibinfo {author} {\bibfnamefont {A.~J.}\
  \bibnamefont {Mundt}}, \bibinfo {author} {\bibfnamefont {C.~J.}\ \bibnamefont
  {Kaine}}, \bibinfo {author} {\bibfnamefont {B.~S.}\ \bibnamefont {Carter}},
  \bibinfo {author} {\bibfnamefont {W.~G.}\ \bibnamefont {Bradley}}, \ and\
  \bibinfo {author} {\bibfnamefont {A.~M.}\ \bibnamefont {Dale}},\ }\href
  {\doibase 10.1158/0008-5472.CAN-13-3534} {\bibfield  {journal} {\bibinfo
  {journal} {Cancer Res.}\ }\textbf {\bibinfo {volume} {74}},\ \bibinfo {pages}
  {4638} (\bibinfo {year} {2014})}\BibitemShut {NoStop}%
\bibitem [{\citenamefont {Enzinger}\ \emph {et~al.}(2015)\citenamefont
  {Enzinger}, \citenamefont {Barkhof}, \citenamefont {Ciccarelli},
  \citenamefont {Filippi}, \citenamefont {Kappos}, \citenamefont {Rocca},
  \citenamefont {Ropele}, \citenamefont {Rovira}, \citenamefont {Schneider},
  \citenamefont {de~Stefano}, \citenamefont {Vrenken}, \citenamefont
  {Wheeler-Kingshott}, \citenamefont {Wuerfel}, \citenamefont {Fazekas},\ and\
  \citenamefont {and}}]{onbehalfoftheMAGNIMSstudygroup2015}%
  \BibitemOpen
  \bibfield  {author} {\bibinfo {author} {\bibfnamefont {C.}~\bibnamefont
  {Enzinger}}, \bibinfo {author} {\bibfnamefont {F.}~\bibnamefont {Barkhof}},
  \bibinfo {author} {\bibfnamefont {O.}~\bibnamefont {Ciccarelli}}, \bibinfo
  {author} {\bibfnamefont {M.}~\bibnamefont {Filippi}}, \bibinfo {author}
  {\bibfnamefont {L.}~\bibnamefont {Kappos}}, \bibinfo {author} {\bibfnamefont
  {M.~A.}\ \bibnamefont {Rocca}}, \bibinfo {author} {\bibfnamefont
  {S.}~\bibnamefont {Ropele}}, \bibinfo {author} {\bibfnamefont
  {{\`A}.}~\bibnamefont {Rovira}}, \bibinfo {author} {\bibfnamefont
  {T.}~\bibnamefont {Schneider}}, \bibinfo {author} {\bibfnamefont
  {N.}~\bibnamefont {de~Stefano}}, \bibinfo {author} {\bibfnamefont
  {H.}~\bibnamefont {Vrenken}}, \bibinfo {author} {\bibfnamefont
  {C.}~\bibnamefont {Wheeler-Kingshott}}, \bibinfo {author} {\bibfnamefont
  {J.}~\bibnamefont {Wuerfel}}, \bibinfo {author} {\bibfnamefont
  {F.}~\bibnamefont {Fazekas}}, \ and\ \bibinfo {author} {\bibfnamefont
  {o.}~\bibnamefont {and}},\ }\href {\doibase 10.1038/nrneurol.2015.194}
  {\bibfield  {journal} {\bibinfo  {journal} {Nat. Rev. Neurol.}\ }\textbf
  {\bibinfo {volume} {11}},\ \bibinfo {pages} {676} (\bibinfo {year}
  {2015})}\BibitemShut {NoStop}%
\bibitem [{\citenamefont {Hursh}(1939)}]{Hursh1939}%
  \BibitemOpen
  \bibfield  {author} {\bibinfo {author} {\bibfnamefont {J.~B.}\ \bibnamefont
  {Hursh}},\ }\href {\doibase 10.1152/ajplegacy.1939.127.1.131} {\bibfield
  {journal} {\bibinfo  {journal} {Am. J. Physiol.}\ }\textbf {\bibinfo {volume}
  {127}},\ \bibinfo {pages} {131} (\bibinfo {year} {1939})}\BibitemShut
  {NoStop}%
\bibitem [{\citenamefont {Waxman}\ and\ \citenamefont
  {Bennett}(1972)}]{Waxman1972}%
  \BibitemOpen
  \bibfield  {author} {\bibinfo {author} {\bibfnamefont {S.~G.}\ \bibnamefont
  {Waxman}}\ and\ \bibinfo {author} {\bibfnamefont {M.~V.~L.}\ \bibnamefont
  {Bennett}},\ }\href {\doibase 10.1038/newbio238217a0} {\bibfield  {journal}
  {\bibinfo  {journal} {Nature New Biol.}\ }\textbf {\bibinfo {volume} {238}},\
  \bibinfo {pages} {217} (\bibinfo {year} {1972})}\BibitemShut {NoStop}%
\bibitem [{\citenamefont {Drago}\ \emph {et~al.}(2011)\citenamefont {Drago},
  \citenamefont {Babiloni}, \citenamefont {Bartr{\'e}s-Faz}, \citenamefont
  {Caroli}, \citenamefont {Bosch}, \citenamefont {Hensch}, \citenamefont
  {Didic}, \citenamefont {Klafki}, \citenamefont {Pievani}, \citenamefont
  {Jovicich}, \citenamefont {Venturi}, \citenamefont {Spitzer}, \citenamefont
  {Vecchio}, \citenamefont {Schoenknecht}, \citenamefont {Wiltfang},
  \citenamefont {Redolfi}, \citenamefont {Forloni}, \citenamefont {Blin},
  \citenamefont {Irving}, \citenamefont {Davis}, \citenamefont {H{\r
  a}rdemark},\ and\ \citenamefont {Frisoni}}]{Drago2011}%
  \BibitemOpen
  \bibfield  {author} {\bibinfo {author} {\bibfnamefont {V.}~\bibnamefont
  {Drago}}, \bibinfo {author} {\bibfnamefont {C.}~\bibnamefont {Babiloni}},
  \bibinfo {author} {\bibfnamefont {D.}~\bibnamefont {Bartr{\'e}s-Faz}},
  \bibinfo {author} {\bibfnamefont {A.}~\bibnamefont {Caroli}}, \bibinfo
  {author} {\bibfnamefont {B.}~\bibnamefont {Bosch}}, \bibinfo {author}
  {\bibfnamefont {T.}~\bibnamefont {Hensch}}, \bibinfo {author} {\bibfnamefont
  {M.}~\bibnamefont {Didic}}, \bibinfo {author} {\bibfnamefont {H.-W.}\
  \bibnamefont {Klafki}}, \bibinfo {author} {\bibfnamefont {M.}~\bibnamefont
  {Pievani}}, \bibinfo {author} {\bibfnamefont {J.}~\bibnamefont {Jovicich}},
  \bibinfo {author} {\bibfnamefont {L.}~\bibnamefont {Venturi}}, \bibinfo
  {author} {\bibfnamefont {P.}~\bibnamefont {Spitzer}}, \bibinfo {author}
  {\bibfnamefont {F.}~\bibnamefont {Vecchio}}, \bibinfo {author} {\bibfnamefont
  {P.}~\bibnamefont {Schoenknecht}}, \bibinfo {author} {\bibfnamefont
  {J.}~\bibnamefont {Wiltfang}}, \bibinfo {author} {\bibfnamefont
  {A.}~\bibnamefont {Redolfi}}, \bibinfo {author} {\bibfnamefont
  {G.}~\bibnamefont {Forloni}}, \bibinfo {author} {\bibfnamefont
  {O.}~\bibnamefont {Blin}}, \bibinfo {author} {\bibfnamefont {E.}~\bibnamefont
  {Irving}}, \bibinfo {author} {\bibfnamefont {C.}~\bibnamefont {Davis}},
  \bibinfo {author} {\bibfnamefont {H.-g.}\ \bibnamefont {H{\r a}rdemark}}, \
  and\ \bibinfo {author} {\bibfnamefont {G.~B.}\ \bibnamefont {Frisoni}},\
  }\href {\doibase 10.3233/JAD-2011-0043} {\bibfield  {journal} {\bibinfo
  {journal} {J. Alzheimers Dis.}\ }\textbf {\bibinfo {volume} {26}},\ \bibinfo
  {pages} {159} (\bibinfo {year} {2011})}\BibitemShut {NoStop}%
\bibitem [{\citenamefont {White}\ \emph {et~al.}(2013)\citenamefont {White},
  \citenamefont {Leergaard}, \citenamefont {D'Arceuil}, \citenamefont
  {Bjaalie},\ and\ \citenamefont {Dale}}]{White2013}%
  \BibitemOpen
  \bibfield  {author} {\bibinfo {author} {\bibfnamefont {N.~S.}\ \bibnamefont
  {White}}, \bibinfo {author} {\bibfnamefont {T.~B.}\ \bibnamefont
  {Leergaard}}, \bibinfo {author} {\bibfnamefont {H.}~\bibnamefont
  {D'Arceuil}}, \bibinfo {author} {\bibfnamefont {J.~G.}\ \bibnamefont
  {Bjaalie}}, \ and\ \bibinfo {author} {\bibfnamefont {A.~M.}\ \bibnamefont
  {Dale}},\ }\href {\doibase 10.1002/hbm.21454} {\bibfield  {journal} {\bibinfo
   {journal} {Hum. Brain Mapp.}\ }\textbf {\bibinfo {volume} {34}},\ \bibinfo
  {pages} {327} (\bibinfo {year} {2013})}\BibitemShut {NoStop}%
\bibitem [{\citenamefont {Xu}\ \emph {et~al.}(2014)\citenamefont {Xu},
  \citenamefont {Li}, \citenamefont {Harkins}, \citenamefont {Jiang},
  \citenamefont {Xie}, \citenamefont {Kang}, \citenamefont {Does},\ and\
  \citenamefont {Gore}}]{Xu2014}%
  \BibitemOpen
  \bibfield  {author} {\bibinfo {author} {\bibfnamefont {J.}~\bibnamefont
  {Xu}}, \bibinfo {author} {\bibfnamefont {H.}~\bibnamefont {Li}}, \bibinfo
  {author} {\bibfnamefont {K.~D.}\ \bibnamefont {Harkins}}, \bibinfo {author}
  {\bibfnamefont {X.}~\bibnamefont {Jiang}}, \bibinfo {author} {\bibfnamefont
  {J.}~\bibnamefont {Xie}}, \bibinfo {author} {\bibfnamefont {H.}~\bibnamefont
  {Kang}}, \bibinfo {author} {\bibfnamefont {M.~D.}\ \bibnamefont {Does}}, \
  and\ \bibinfo {author} {\bibfnamefont {J.~C.}\ \bibnamefont {Gore}},\ }\href
  {\doibase 10.1016/j.neuroimage.2014.09.006} {\bibfield  {journal} {\bibinfo
  {journal} {Neuroimage}\ }\textbf {\bibinfo {volume} {103}},\ \bibinfo {pages}
  {10} (\bibinfo {year} {2014})}\BibitemShut {NoStop}%
\bibitem [{\citenamefont {Grussu}\ \emph {et~al.}(2017)\citenamefont {Grussu},
  \citenamefont {Schneider}, \citenamefont {Tur}, \citenamefont {Yates},
  \citenamefont {Tachrount}, \citenamefont {Ianu{\c s}}, \citenamefont
  {Yiannakas}, \citenamefont {Newcombe}, \citenamefont {Zhang}, \citenamefont
  {Alexander}, \citenamefont {DeLuca},\ and\ \citenamefont {Gandini
  Wheeler-Kingshott}}]{Grussu2017}%
  \BibitemOpen
  \bibfield  {author} {\bibinfo {author} {\bibfnamefont {F.}~\bibnamefont
  {Grussu}}, \bibinfo {author} {\bibfnamefont {T.}~\bibnamefont {Schneider}},
  \bibinfo {author} {\bibfnamefont {C.}~\bibnamefont {Tur}}, \bibinfo {author}
  {\bibfnamefont {R.~L.}\ \bibnamefont {Yates}}, \bibinfo {author}
  {\bibfnamefont {M.}~\bibnamefont {Tachrount}}, \bibinfo {author}
  {\bibfnamefont {A.}~\bibnamefont {Ianu{\c s}}}, \bibinfo {author}
  {\bibfnamefont {M.~C.}\ \bibnamefont {Yiannakas}}, \bibinfo {author}
  {\bibfnamefont {J.}~\bibnamefont {Newcombe}}, \bibinfo {author}
  {\bibfnamefont {H.}~\bibnamefont {Zhang}}, \bibinfo {author} {\bibfnamefont
  {D.~C.}\ \bibnamefont {Alexander}}, \bibinfo {author} {\bibfnamefont {G.~C.}\
  \bibnamefont {DeLuca}}, \ and\ \bibinfo {author} {\bibfnamefont {C.~A.~M.}\
  \bibnamefont {Gandini Wheeler-Kingshott}},\ }\href {\doibase
  10.1002/acn3.445} {\bibfield  {journal} {\bibinfo  {journal} {Ann. Clin.
  Transl. Neurol.}\ }\textbf {\bibinfo {volume} {4}},\ \bibinfo {pages} {663}
  (\bibinfo {year} {2017})}\BibitemShut {NoStop}%
\bibitem [{\citenamefont {Assaf}\ \emph {et~al.}(2008)\citenamefont {Assaf},
  \citenamefont {Blumenfeld-Katzir}, \citenamefont {Yovel},\ and\ \citenamefont
  {Basser}}]{Assaf2008}%
  \BibitemOpen
  \bibfield  {author} {\bibinfo {author} {\bibfnamefont {Y.}~\bibnamefont
  {Assaf}}, \bibinfo {author} {\bibfnamefont {T.}~\bibnamefont
  {Blumenfeld-Katzir}}, \bibinfo {author} {\bibfnamefont {Y.}~\bibnamefont
  {Yovel}}, \ and\ \bibinfo {author} {\bibfnamefont {P.~J.}\ \bibnamefont
  {Basser}},\ }\href {\doibase 10.1002/mrm.21577} {\bibfield  {journal}
  {\bibinfo  {journal} {Magn. Reson. Med.}\ }\textbf {\bibinfo {volume} {59}},\
  \bibinfo {pages} {1347} (\bibinfo {year} {2008})},\ \Eprint
  {http://arxiv.org/abs/https://onlinelibrary.wiley.com/doi/pdf/10.1002/mrm.21577}
  {https://onlinelibrary.wiley.com/doi/pdf/10.1002/mrm.21577} \BibitemShut
  {NoStop}%
\bibitem [{\citenamefont {Alexander}\ \emph {et~al.}(2010)\citenamefont
  {Alexander}, \citenamefont {Hubbard}, \citenamefont {Hall}, \citenamefont
  {Moore}, \citenamefont {Ptito}, \citenamefont {Parker},\ and\ \citenamefont
  {Dyrby}}]{Alexander2010}%
  \BibitemOpen
  \bibfield  {author} {\bibinfo {author} {\bibfnamefont {D.~C.}\ \bibnamefont
  {Alexander}}, \bibinfo {author} {\bibfnamefont {P.~L.}\ \bibnamefont
  {Hubbard}}, \bibinfo {author} {\bibfnamefont {M.~G.}\ \bibnamefont {Hall}},
  \bibinfo {author} {\bibfnamefont {E.~A.}\ \bibnamefont {Moore}}, \bibinfo
  {author} {\bibfnamefont {M.}~\bibnamefont {Ptito}}, \bibinfo {author}
  {\bibfnamefont {G.~J.}\ \bibnamefont {Parker}}, \ and\ \bibinfo {author}
  {\bibfnamefont {T.~B.}\ \bibnamefont {Dyrby}},\ }\href {\doibase
  https://doi.org/10.1016/j.neuroimage.2010.05.043} {\bibfield  {journal}
  {\bibinfo  {journal} {Neuroimage}\ }\textbf {\bibinfo {volume} {52}},\
  \bibinfo {pages} {1374 } (\bibinfo {year} {2010})}\BibitemShut {NoStop}%
\bibitem [{\citenamefont {Shemesh}\ \emph {et~al.}(2015)\citenamefont
  {Shemesh}, \citenamefont {\'Alvarez},\ and\ \citenamefont
  {Frydman}}]{Shemesh2015}%
  \BibitemOpen
  \bibfield  {author} {\bibinfo {author} {\bibfnamefont {N.}~\bibnamefont
  {Shemesh}}, \bibinfo {author} {\bibfnamefont {G.~A.}\ \bibnamefont
  {\'Alvarez}}, \ and\ \bibinfo {author} {\bibfnamefont {L.}~\bibnamefont
  {Frydman}},\ }\href {\doibase 10.1371/journal.pone.0133201} {\bibfield
  {journal} {\bibinfo  {journal} {PLoS One}\ }\textbf {\bibinfo {volume}
  {10}},\ \bibinfo {pages} {e0133201} (\bibinfo {year} {2015})}\BibitemShut
  {NoStop}%
\bibitem [{\citenamefont {Lambert}\ \emph {et~al.}(2009)\citenamefont
  {Lambert}, \citenamefont {Hergenr{\"o}der}, \citenamefont {Suter},\ and\
  \citenamefont {Deckert}}]{Lambert:2009kx}%
  \BibitemOpen
  \bibfield  {author} {\bibinfo {author} {\bibfnamefont {J.}~\bibnamefont
  {Lambert}}, \bibinfo {author} {\bibfnamefont {R.}~\bibnamefont
  {Hergenr{\"o}der}}, \bibinfo {author} {\bibfnamefont {D.}~\bibnamefont
  {Suter}}, \ and\ \bibinfo {author} {\bibfnamefont {V.}~\bibnamefont
  {Deckert}},\ }\href@noop {} {\bibfield  {journal} {\bibinfo  {journal}
  {Angew. Chem. Int. Ed.}\ }\textbf {\bibinfo {volume} {48}},\ \bibinfo {pages}
  {6343} (\bibinfo {year} {2009})}\BibitemShut {NoStop}%
\bibitem [{\citenamefont {Moore}\ and\ \citenamefont
  {Tycko}(2015)}]{MOORE20151}%
  \BibitemOpen
  \bibfield  {author} {\bibinfo {author} {\bibfnamefont {E.}~\bibnamefont
  {Moore}}\ and\ \bibinfo {author} {\bibfnamefont {R.}~\bibnamefont {Tycko}},\
  }\href {\doibase http://dx.doi.org/10.1016/j.jmr.2015.09.001} {\bibfield
  {journal} {\bibinfo  {journal} {J. Magn. Reson.}\ }\textbf {\bibinfo {volume}
  {260}},\ \bibinfo {pages} {1 } (\bibinfo {year} {2015})}\BibitemShut
  {NoStop}%
\bibitem [{\citenamefont {Stejskal}\ and\ \citenamefont
  {Tanner}(1965)}]{Stejskal1965}%
  \BibitemOpen
  \bibfield  {author} {\bibinfo {author} {\bibfnamefont {E.~O.}\ \bibnamefont
  {Stejskal}}\ and\ \bibinfo {author} {\bibfnamefont {J.~E.}\ \bibnamefont
  {Tanner}},\ }\href {\doibase 10.1063/1.1695690} {\bibfield  {journal}
  {\bibinfo  {journal} {J. Chem. Phys.}\ }\textbf {\bibinfo {volume} {42}},\
  \bibinfo {pages} {288} (\bibinfo {year} {1965})},\ \Eprint
  {http://arxiv.org/abs/https://doi.org/10.1063/1.1695690}
  {https://doi.org/10.1063/1.1695690} \BibitemShut {NoStop}%
\bibitem [{\citenamefont {Le~Bihan}(2003)}]{LeBihan2003}%
  \BibitemOpen
  \bibfield  {author} {\bibinfo {author} {\bibfnamefont {D.}~\bibnamefont
  {Le~Bihan}},\ }\href {\doibase 10.1038/nrn1119} {\bibfield  {journal}
  {\bibinfo  {journal} {Nat. Rev. Neurosci.}\ }\textbf {\bibinfo {volume}
  {4}},\ \bibinfo {pages} {469} (\bibinfo {year} {2003})}\BibitemShut {NoStop}%
\bibitem [{\citenamefont {Grebenkov}(2007)}]{Grebenkov2007}%
  \BibitemOpen
  \bibfield  {author} {\bibinfo {author} {\bibfnamefont {D.~S.}\ \bibnamefont
  {Grebenkov}},\ }\href {\doibase 10.1103/RevModPhys.79.1077} {\bibfield
  {journal} {\bibinfo  {journal} {Rev. Mod. Phys.}\ }\textbf {\bibinfo {volume}
  {79}},\ \bibinfo {pages} {1077} (\bibinfo {year} {2007})}\BibitemShut
  {NoStop}%
\bibitem [{\citenamefont {Callaghan}(2011)}]{Callaghan2011}%
  \BibitemOpen
  \bibfield  {author} {\bibinfo {author} {\bibfnamefont {P.~T.}\ \bibnamefont
  {Callaghan}},\ }\href@noop {} {\emph {\bibinfo {title} {Translational
  Dynamics and Magnetic {Resonance:Principles} of Pulsed Gradient Spin Echo
  {NMR}}}}\ (\bibinfo  {publisher} {Oxford University Press},\ \bibinfo
  {address} {Oxford},\ \bibinfo {year} {2011})\BibitemShut {NoStop}%
\bibitem [{\citenamefont {Stepisnik}(1993)}]{Stepisnik1993}%
  \BibitemOpen
  \bibfield  {author} {\bibinfo {author} {\bibfnamefont {J.}~\bibnamefont
  {Stepisnik}},\ }\href {\doibase 10.1016/0921-4526(93)90124-O} {\bibfield
  {journal} {\bibinfo  {journal} {Physica B}\ }\textbf {\bibinfo {volume}
  {183}},\ \bibinfo {pages} {343} (\bibinfo {year} {1993})}\BibitemShut
  {NoStop}%
\bibitem [{\citenamefont {Callaghan}\ and\ \citenamefont
  {Stepisnik}(1995)}]{Callaghan1995}%
  \BibitemOpen
  \bibfield  {author} {\bibinfo {author} {\bibfnamefont {P.~T.}\ \bibnamefont
  {Callaghan}}\ and\ \bibinfo {author} {\bibfnamefont {J.}~\bibnamefont
  {Stepisnik}},\ }\href
  {http://citeseerx.ist.psu.edu/viewdoc/download?doi=10.1.1.159.4333&rep=rep1&type=pdf}
  {\bibfield  {journal} {\bibinfo  {journal} {J. Magn. Reson.}\ }\textbf
  {\bibinfo {volume} {117}},\ \bibinfo {pages} {118} (\bibinfo {year}
  {1995})}\BibitemShut {NoStop}%
\bibitem [{\citenamefont {Callaghan}(1997)}]{Callaghan1997}%
  \BibitemOpen
  \bibfield  {author} {\bibinfo {author} {\bibfnamefont {P.~T.}\ \bibnamefont
  {Callaghan}},\ }\href {\doibase https://doi.org/10.1006/jmre.1997.1233}
  {\bibfield  {journal} {\bibinfo  {journal} {J. Magn. Reson.}\ }\textbf
  {\bibinfo {volume} {129}},\ \bibinfo {pages} {74 } (\bibinfo {year}
  {1997})}\BibitemShut {NoStop}%
\bibitem [{\citenamefont {Shemesh}\ \emph {et~al.}(2013)\citenamefont
  {Shemesh}, \citenamefont {\'Alvarez},\ and\ \citenamefont
  {Frydman}}]{Shemesh2013}%
  \BibitemOpen
  \bibfield  {author} {\bibinfo {author} {\bibfnamefont {N.}~\bibnamefont
  {Shemesh}}, \bibinfo {author} {\bibfnamefont {G.~A.}\ \bibnamefont
  {\'Alvarez}}, \ and\ \bibinfo {author} {\bibfnamefont {L.}~\bibnamefont
  {Frydman}},\ }\href
  {http://www.sciencedirect.com/science/article/pii/S1090780713002309}
  {\bibfield  {journal} {\bibinfo  {journal} {J. Magn. Reson.}\ }\textbf
  {\bibinfo {volume} {237}},\ \bibinfo {pages} {49 } (\bibinfo {year}
  {2013})}\BibitemShut {NoStop}%
\bibitem [{\citenamefont {Drobnjak}\ \emph {et~al.}(2016)\citenamefont
  {Drobnjak}, \citenamefont {Zhang}, \citenamefont {Ianus}, \citenamefont
  {Kaden},\ and\ \citenamefont {Alexander}}]{Drobnjak2016}%
  \BibitemOpen
  \bibfield  {author} {\bibinfo {author} {\bibfnamefont {I.}~\bibnamefont
  {Drobnjak}}, \bibinfo {author} {\bibfnamefont {H.}~\bibnamefont {Zhang}},
  \bibinfo {author} {\bibfnamefont {A.}~\bibnamefont {Ianus}}, \bibinfo
  {author} {\bibfnamefont {E.}~\bibnamefont {Kaden}}, \ and\ \bibinfo {author}
  {\bibfnamefont {D.~C.}\ \bibnamefont {Alexander}},\ }\href {\doibase
  10.1002/mrm.25631} {\bibfield  {journal} {\bibinfo  {journal} {Magn. Reson.
  Med.}\ }\textbf {\bibinfo {volume} {75}},\ \bibinfo {pages} {688} (\bibinfo
  {year} {2016})},\ \Eprint
  {http://arxiv.org/abs/https://onlinelibrary.wiley.com/doi/pdf/10.1002/mrm.25631}
  {https://onlinelibrary.wiley.com/doi/pdf/10.1002/mrm.25631} \BibitemShut
  {NoStop}%
\bibitem [{\citenamefont {Nilsson}\ \emph {et~al.}(2017)\citenamefont
  {Nilsson}, \citenamefont {Lasic}, \citenamefont {Drobnjak}, \citenamefont
  {Topgaard},\ and\ \citenamefont {Westin}}]{Nilsson2017}%
  \BibitemOpen
  \bibfield  {author} {\bibinfo {author} {\bibfnamefont {M.}~\bibnamefont
  {Nilsson}}, \bibinfo {author} {\bibfnamefont {S.}~\bibnamefont {Lasic}},
  \bibinfo {author} {\bibfnamefont {I.}~\bibnamefont {Drobnjak}}, \bibinfo
  {author} {\bibfnamefont {D.}~\bibnamefont {Topgaard}}, \ and\ \bibinfo
  {author} {\bibfnamefont {C.-F.}\ \bibnamefont {Westin}},\ }\href {\doibase
  10.1002/nbm.3711} {\bibfield  {journal} {\bibinfo  {journal} {NMR Biomed.}\
  }\textbf {\bibinfo {volume} {30}},\ \bibinfo {pages} {e3711} (\bibinfo {year}
  {2017})},\ \bibinfo {note} {e3711 nbm.3711},\ \Eprint
  {http://arxiv.org/abs/https://onlinelibrary.wiley.com/doi/pdf/10.1002/nbm.3711}
  {https://onlinelibrary.wiley.com/doi/pdf/10.1002/nbm.3711} \BibitemShut
  {NoStop}%
\bibitem [{\citenamefont {Kakkar}\ \emph {et~al.}(2018)\citenamefont {Kakkar},
  \citenamefont {Bennett}, \citenamefont {Siow}, \citenamefont {Richardson},
  \citenamefont {Ianus}, \citenamefont {Quick}, \citenamefont {Atkinson},
  \citenamefont {Phillips},\ and\ \citenamefont {Drobnjak}}]{Kakkar2018}%
  \BibitemOpen
  \bibfield  {author} {\bibinfo {author} {\bibfnamefont {L.~S.}\ \bibnamefont
  {Kakkar}}, \bibinfo {author} {\bibfnamefont {O.~F.}\ \bibnamefont {Bennett}},
  \bibinfo {author} {\bibfnamefont {B.}~\bibnamefont {Siow}}, \bibinfo {author}
  {\bibfnamefont {S.}~\bibnamefont {Richardson}}, \bibinfo {author}
  {\bibfnamefont {A.}~\bibnamefont {Ianus}}, \bibinfo {author} {\bibfnamefont
  {T.}~\bibnamefont {Quick}}, \bibinfo {author} {\bibfnamefont
  {D.}~\bibnamefont {Atkinson}}, \bibinfo {author} {\bibfnamefont {J.~B.}\
  \bibnamefont {Phillips}}, \ and\ \bibinfo {author} {\bibfnamefont
  {I.}~\bibnamefont {Drobnjak}},\ }\href {\doibase
  https://doi.org/10.1016/j.neuroimage.2017.07.060} {\bibfield  {journal}
  {\bibinfo  {journal} {Neuroimage}\ }\textbf {\bibinfo {volume} {182}},\
  \bibinfo {pages} {314 } (\bibinfo {year} {2018})},\ \bibinfo {note}
  {microstructural Imaging}\BibitemShut {NoStop}%
\bibitem [{\citenamefont {Ong}\ and\ \citenamefont {Wehrli}(2010)}]{Ong2010}%
  \BibitemOpen
  \bibfield  {author} {\bibinfo {author} {\bibfnamefont {H.~H.}\ \bibnamefont
  {Ong}}\ and\ \bibinfo {author} {\bibfnamefont {F.~W.}\ \bibnamefont
  {Wehrli}},\ }\href {\doibase
  https://doi.org/10.1016/j.neuroimage.2010.03.063} {\bibfield  {journal}
  {\bibinfo  {journal} {Neuroimage}\ }\textbf {\bibinfo {volume} {51}},\
  \bibinfo {pages} {1360 } (\bibinfo {year} {2010})}\BibitemShut {NoStop}%
\bibitem [{\citenamefont {Komlosh}\ \emph {et~al.}(2011)\citenamefont
  {Komlosh}, \citenamefont {Ozarslan}, \citenamefont {Lizak}, \citenamefont
  {Horkay}, \citenamefont {Schram}, \citenamefont {Shemesh}, \citenamefont
  {Cohen},\ and\ \citenamefont {Basser}}]{Komlosh2011}%
  \BibitemOpen
  \bibfield  {author} {\bibinfo {author} {\bibfnamefont {M.~E.}\ \bibnamefont
  {Komlosh}}, \bibinfo {author} {\bibfnamefont {E.}~\bibnamefont {Ozarslan}},
  \bibinfo {author} {\bibfnamefont {M.~J.}\ \bibnamefont {Lizak}}, \bibinfo
  {author} {\bibfnamefont {F.}~\bibnamefont {Horkay}}, \bibinfo {author}
  {\bibfnamefont {V.}~\bibnamefont {Schram}}, \bibinfo {author} {\bibfnamefont
  {N.}~\bibnamefont {Shemesh}}, \bibinfo {author} {\bibfnamefont
  {Y.}~\bibnamefont {Cohen}}, \ and\ \bibinfo {author} {\bibfnamefont {P.~J.}\
  \bibnamefont {Basser}},\ }\href {\doibase
  https://doi.org/10.1016/j.jmr.2010.10.014} {\bibfield  {journal} {\bibinfo
  {journal} {J. Magn. Reson.}\ }\textbf {\bibinfo {volume} {208}},\ \bibinfo
  {pages} {128 } (\bibinfo {year} {2011})}\BibitemShut {NoStop}%
\bibitem [{\citenamefont {\'Alvarez}\ \emph {et~al.}(2013)\citenamefont
  {\'Alvarez}, \citenamefont {Shemesh},\ and\ \citenamefont
  {Frydman}}]{Alvarez2013a}%
  \BibitemOpen
  \bibfield  {author} {\bibinfo {author} {\bibfnamefont {G.~A.}\ \bibnamefont
  {\'Alvarez}}, \bibinfo {author} {\bibfnamefont {N.}~\bibnamefont {Shemesh}},
  \ and\ \bibinfo {author} {\bibfnamefont {L.}~\bibnamefont {Frydman}},\ }\href
  {\doibase 10.1103/PhysRevLett.111.080404} {\bibfield  {journal} {\bibinfo
  {journal} {Phys. Rev. Lett.}\ }\textbf {\bibinfo {volume} {111}},\ \bibinfo
  {pages} {080404} (\bibinfo {year} {2013})}\BibitemShut {NoStop}%
\bibitem [{\citenamefont {Xu}\ \emph {et~al.}(2019)\citenamefont {Xu},
  \citenamefont {Jiang}, \citenamefont {Li}, \citenamefont {Arlinghaus},
  \citenamefont {McKinley}, \citenamefont {Devan}, \citenamefont {Hardy},
  \citenamefont {Xie}, \citenamefont {Kang}, \citenamefont {Chakravarthy},\
  and\ \citenamefont {Gore}}]{Xu2019}%
  \BibitemOpen
  \bibfield  {author} {\bibinfo {author} {\bibfnamefont {J.}~\bibnamefont
  {Xu}}, \bibinfo {author} {\bibfnamefont {X.}~\bibnamefont {Jiang}}, \bibinfo
  {author} {\bibfnamefont {H.}~\bibnamefont {Li}}, \bibinfo {author}
  {\bibfnamefont {L.~R.}\ \bibnamefont {Arlinghaus}}, \bibinfo {author}
  {\bibfnamefont {E.~T.}\ \bibnamefont {McKinley}}, \bibinfo {author}
  {\bibfnamefont {S.~P.}\ \bibnamefont {Devan}}, \bibinfo {author}
  {\bibfnamefont {B.~M.}\ \bibnamefont {Hardy}}, \bibinfo {author}
  {\bibfnamefont {J.}~\bibnamefont {Xie}}, \bibinfo {author} {\bibfnamefont
  {H.}~\bibnamefont {Kang}}, \bibinfo {author} {\bibfnamefont {A.~B.}\
  \bibnamefont {Chakravarthy}}, \ and\ \bibinfo {author} {\bibfnamefont
  {J.~C.}\ \bibnamefont {Gore}},\ }\href {\doibase 10.1002/mrm.28056}
  {\bibfield  {journal} {\bibinfo  {journal} {Magn. Reson. Med.}\ ,\ \bibinfo
  {pages} {mrm.28056}} (\bibinfo {year} {2019})}\BibitemShut {NoStop}%
\bibitem [{\citenamefont {Stepisnik}(1999)}]{Stepisnik1999}%
  \BibitemOpen
  \bibfield  {author} {\bibinfo {author} {\bibfnamefont {J.}~\bibnamefont
  {Stepisnik}},\ }\href {\doibase 10.1016/S0921-4526(99)00160-X} {\bibfield
  {journal} {\bibinfo  {journal} {Physica B}\ }\textbf {\bibinfo {volume}
  {270}},\ \bibinfo {pages} {110} (\bibinfo {year} {1999})}\BibitemShut
  {NoStop}%
\bibitem [{\citenamefont {Stepisnik}\ \emph {et~al.}(2006)\citenamefont
  {Stepisnik}, \citenamefont {Lasic}, \citenamefont {Mohorix}, \citenamefont
  {Sersa},\ and\ \citenamefont {Sepe}}]{Stepisnik2006}%
  \BibitemOpen
  \bibfield  {author} {\bibinfo {author} {\bibfnamefont {J.}~\bibnamefont
  {Stepisnik}}, \bibinfo {author} {\bibfnamefont {S.}~\bibnamefont {Lasic}},
  \bibinfo {author} {\bibfnamefont {A.}~\bibnamefont {Mohorix}}, \bibinfo
  {author} {\bibfnamefont {I.}~\bibnamefont {Sersa}}, \ and\ \bibinfo {author}
  {\bibfnamefont {A.}~\bibnamefont {Sepe}},\ }\href {\doibase
  10.1016/j.jmr.2006.06.023} {\bibfield  {journal} {\bibinfo  {journal} {J.
  Magn. Reson.}\ }\textbf {\bibinfo {volume} {182}},\ \bibinfo {pages} {195}
  (\bibinfo {year} {2006})}\BibitemShut {NoStop}%
\bibitem [{\citenamefont {Lasic}\ \emph {et~al.}(2006)\citenamefont {Lasic},
  \citenamefont {Stepisnik},\ and\ \citenamefont {Mohoric}}]{Lasic2006}%
  \BibitemOpen
  \bibfield  {author} {\bibinfo {author} {\bibfnamefont {S.}~\bibnamefont
  {Lasic}}, \bibinfo {author} {\bibfnamefont {J.}~\bibnamefont {Stepisnik}}, \
  and\ \bibinfo {author} {\bibfnamefont {A.}~\bibnamefont {Mohoric}},\ }\href
  {\doibase 10.1016/j.jmr.2006.06.030} {\bibfield  {journal} {\bibinfo
  {journal} {J. Magn. Reson.}\ }\textbf {\bibinfo {volume} {182}},\ \bibinfo
  {pages} {208} (\bibinfo {year} {2006})}\BibitemShut {NoStop}%
\bibitem [{\citenamefont {\'Alvarez}\ and\ \citenamefont
  {Suter}(2011)}]{Alvarez2011a}%
  \BibitemOpen
  \bibfield  {author} {\bibinfo {author} {\bibfnamefont {G.~A.}\ \bibnamefont
  {\'Alvarez}}\ and\ \bibinfo {author} {\bibfnamefont {D.}~\bibnamefont
  {Suter}},\ }\href {\doibase 10.1103/PhysRevLett.107.230501} {\bibfield
  {journal} {\bibinfo  {journal} {Phys. Rev. Lett.}\ }\textbf {\bibinfo
  {volume} {107}},\ \bibinfo {pages} {230501} (\bibinfo {year}
  {2011})}\BibitemShut {NoStop}%
\bibitem [{\citenamefont {Kofman}\ and\ \citenamefont
  {Kurizki}(2000)}]{kofman2000acceleration}%
  \BibitemOpen
  \bibfield  {author} {\bibinfo {author} {\bibfnamefont {A.}~\bibnamefont
  {Kofman}}\ and\ \bibinfo {author} {\bibfnamefont {G.}~\bibnamefont
  {Kurizki}},\ }\href@noop {} {\bibfield  {journal} {\bibinfo  {journal}
  {Nature}\ }\textbf {\bibinfo {volume} {405}},\ \bibinfo {pages} {546}
  (\bibinfo {year} {2000})}\BibitemShut {NoStop}%
\bibitem [{\citenamefont {Kofman}\ and\ \citenamefont
  {Kurizki}(2001)}]{kofman_universal_2001}%
  \BibitemOpen
  \bibfield  {author} {\bibinfo {author} {\bibfnamefont {A.~G.}\ \bibnamefont
  {Kofman}}\ and\ \bibinfo {author} {\bibfnamefont {G.}~\bibnamefont
  {Kurizki}},\ }\href {\doibase 10.1103/PhysRevLett.87.270405} {\bibfield
  {journal} {\bibinfo  {journal} {Phys. Rev. Lett.}\ }\textbf {\bibinfo
  {volume} {87}},\ \bibinfo {pages} {270405} (\bibinfo {year}
  {2001})}\BibitemShut {NoStop}%
\bibitem [{\citenamefont {Kofman}\ and\ \citenamefont
  {Kurizki}(2004)}]{kofman_unified_2004}%
  \BibitemOpen
  \bibfield  {author} {\bibinfo {author} {\bibfnamefont {A.~G.}\ \bibnamefont
  {Kofman}}\ and\ \bibinfo {author} {\bibfnamefont {G.}~\bibnamefont
  {Kurizki}},\ }\href {\doibase 10.1103/PhysRevLett.93.130406} {\bibfield
  {journal} {\bibinfo  {journal} {Phys. Rev. Lett.}\ }\textbf {\bibinfo
  {volume} {93}},\ \bibinfo {pages} {130406} (\bibinfo {year}
  {2004})}\BibitemShut {NoStop}%
\bibitem [{\citenamefont {Gordon}\ \emph {et~al.}(2007)\citenamefont {Gordon},
  \citenamefont {Erez},\ and\ \citenamefont {Kurizki}}]{gordon_universal_2007}%
  \BibitemOpen
  \bibfield  {author} {\bibinfo {author} {\bibfnamefont {G.}~\bibnamefont
  {Gordon}}, \bibinfo {author} {\bibfnamefont {N.}~\bibnamefont {Erez}}, \ and\
  \bibinfo {author} {\bibfnamefont {G.}~\bibnamefont {Kurizki}},\ }\href
  {\doibase 10.1088/0953-4075} {\bibfield  {journal} {\bibinfo  {journal} {J.
  Phys. B: At. Mol. Opt. Phys.}\ }\textbf {\bibinfo {volume} {40}},\ \bibinfo
  {pages} {S75} (\bibinfo {year} {2007})}\BibitemShut {NoStop}%
\bibitem [{\citenamefont {Kurizki}\ and\ \citenamefont
  {Zwick}(2015)}]{Kurizki2015a}%
  \BibitemOpen
  \bibfield  {author} {\bibinfo {author} {\bibfnamefont {G.}~\bibnamefont
  {Kurizki}}\ and\ \bibinfo {author} {\bibfnamefont {A.}~\bibnamefont
  {Zwick}},\ }in\ \href
  {http://eu.wiley.com/WileyCDA/WileyTitle/productCd-1118949692.html} {\emph
  {\bibinfo {booktitle} {"From coherent to incoherent dynamical control of open
  quantum systems", To appear in Adv. Chem. Phys. 159}}},\ \bibinfo {editor}
  {edited by\ \bibinfo {editor} {\bibfnamefont {P.}~\bibnamefont {Brumer}},
  \bibinfo {editor} {\bibfnamefont {S.~A.}\ \bibnamefont {Rice}}, \ and\
  \bibinfo {editor} {\bibfnamefont {A.~R.}\ \bibnamefont {Dinner}}}\ (\bibinfo
  {publisher} {John Wiley},\ \bibinfo {year} {2015})\BibitemShut {NoStop}%
\bibitem [{\citenamefont {Zwick}\ \emph {et~al.}(2016)\citenamefont {Zwick},
  \citenamefont {\'Alvarez},\ and\ \citenamefont {Kurizki}}]{Zwick2016a}%
  \BibitemOpen
  \bibfield  {author} {\bibinfo {author} {\bibfnamefont {A.}~\bibnamefont
  {Zwick}}, \bibinfo {author} {\bibfnamefont {G.~A.}\ \bibnamefont
  {\'Alvarez}}, \ and\ \bibinfo {author} {\bibfnamefont {G.}~\bibnamefont
  {Kurizki}},\ }\href {\doibase 10.1103/PhysRevApplied.5.014007} {\bibfield
  {journal} {\bibinfo  {journal} {Phys. Rev. Applied}\ }\textbf {\bibinfo
  {volume} {5}},\ \bibinfo {pages} {014007} (\bibinfo {year}
  {2016})}\BibitemShut {NoStop}%
\bibitem [{\citenamefont {Klauder}\ and\ \citenamefont
  {Anderson}(1962)}]{Klauder1962}%
  \BibitemOpen
  \bibfield  {author} {\bibinfo {author} {\bibfnamefont {J.~R.}\ \bibnamefont
  {Klauder}}\ and\ \bibinfo {author} {\bibfnamefont {P.~W.}\ \bibnamefont
  {Anderson}},\ }\href {\doibase 10.1103/PhysRev.125.912} {\bibfield  {journal}
  {\bibinfo  {journal} {Phys. Rev.}\ }\textbf {\bibinfo {volume} {125}},\
  \bibinfo {pages} {912} (\bibinfo {year} {1962})}\BibitemShut {NoStop}%
\bibitem [{\citenamefont {Paris}(2009)}]{Paris2009_QUANTUM-ESTIMATION}%
  \BibitemOpen
  \bibfield  {author} {\bibinfo {author} {\bibfnamefont {M.~G.~A.}\
  \bibnamefont {Paris}},\ }\href {\doibase 10.1142/S0219749909004839}
  {\bibfield  {journal} {\bibinfo  {journal} {International Journal of Quantum
  Information}\ }\textbf {\bibinfo {volume} {07}},\ \bibinfo {pages} {125}
  (\bibinfo {year} {2009})}\BibitemShut {NoStop}%
\bibitem [{\citenamefont {Braunstein}\ and\ \citenamefont
  {Caves}(1994)}]{Caves_1994_fisher}%
  \BibitemOpen
  \bibfield  {author} {\bibinfo {author} {\bibfnamefont {S.}~\bibnamefont
  {Braunstein}}\ and\ \bibinfo {author} {\bibfnamefont {C.}~\bibnamefont
  {Caves}},\ }\href {\doibase 10.1103/PhysRevLett.72.3439} {\bibfield
  {journal} {\bibinfo  {journal} {Phys. Rev. Lett.}\ }\textbf {\bibinfo
  {volume} {72}},\ \bibinfo {pages} {3439} (\bibinfo {year}
  {1994})}\BibitemShut {NoStop}%
\bibitem [{\citenamefont {Cram{\'e}r}(1946)}]{cramer1999mathematical}%
  \BibitemOpen
  \bibfield  {author} {\bibinfo {author} {\bibfnamefont {H.}~\bibnamefont
  {Cram{\'e}r}},\ }\href@noop {} {\emph {\bibinfo {title} {Mathematical methods
  of statistics}}}\ (\bibinfo  {publisher} {Princeton university press},\
  \bibinfo {year} {1946})\BibitemShut {NoStop}%
\bibitem [{\citenamefont {Benedetti}\ and\ \citenamefont
  {Paris}(2014)}]{Paris2014_Characterization-of-classical-Gaussian}%
  \BibitemOpen
  \bibfield  {author} {\bibinfo {author} {\bibfnamefont {C.}~\bibnamefont
  {Benedetti}}\ and\ \bibinfo {author} {\bibfnamefont {M.~G.}\ \bibnamefont
  {Paris}},\ }\href@noop {} {\bibfield  {journal} {\bibinfo  {journal} {Phys.
  Lett. A}\ }\textbf {\bibinfo {volume} {378}},\ \bibinfo {pages} {2495}
  (\bibinfo {year} {2014})}\BibitemShut {NoStop}%
\bibitem [{\citenamefont {Kurizki}\ \emph {et~al.}(2015)\citenamefont
  {Kurizki}, \citenamefont {Shahmoon},\ and\ \citenamefont
  {Zwick}}]{Zwick_PhysScrip_2015}%
  \BibitemOpen
  \bibfield  {author} {\bibinfo {author} {\bibfnamefont {G.}~\bibnamefont
  {Kurizki}}, \bibinfo {author} {\bibfnamefont {E.}~\bibnamefont {Shahmoon}}, \
  and\ \bibinfo {author} {\bibfnamefont {A.}~\bibnamefont {Zwick}},\ }\href
  {http://stacks.iop.org/1402-4896/90/i=12/a=128002} {\bibfield  {journal}
  {\bibinfo  {journal} {Phys. Scr.}\ }\textbf {\bibinfo {volume} {90}},\
  \bibinfo {pages} {128002} (\bibinfo {year} {2015})}\BibitemShut {NoStop}%
\bibitem [{\citenamefont {Hahn}(1950)}]{Hahn1950}%
  \BibitemOpen
  \bibfield  {author} {\bibinfo {author} {\bibfnamefont {E.}~\bibnamefont
  {Hahn}},\ }\href {http://link.aps.org/doi/10.1103/PhysRev.80.580} {\bibfield
  {journal} {\bibinfo  {journal} {Phys. Rev.}\ }\textbf {\bibinfo {volume}
  {80}},\ \bibinfo {pages} {580} (\bibinfo {year} {1950})}\BibitemShut
  {NoStop}%
\bibitem [{\citenamefont {Setsompop}\ \emph {et~al.}(2013)\citenamefont
  {Setsompop}, \citenamefont {Kimmlingen}, \citenamefont {Eberlein},
  \citenamefont {Witzel}, \citenamefont {Cohen-Adad}, \citenamefont {McNab},
  \citenamefont {Keil}, \citenamefont {Tisdall}, \citenamefont {Hoecht},
  \citenamefont {Dietz}, \citenamefont {Cauley}, \citenamefont {Tountcheva},
  \citenamefont {Matschl}, \citenamefont {Lenz}, \citenamefont {Heberlein},
  \citenamefont {Potthast}, \citenamefont {Thein}, \citenamefont {Van~Horn},
  \citenamefont {Toga}, \citenamefont {Schmitt}, \citenamefont {Lehne},
  \citenamefont {Rosen}, \citenamefont {Wedeen},\ and\ \citenamefont
  {Wald}}]{Setsompop2013}%
  \BibitemOpen
  \bibfield  {author} {\bibinfo {author} {\bibfnamefont {K.}~\bibnamefont
  {Setsompop}}, \bibinfo {author} {\bibfnamefont {R.}~\bibnamefont
  {Kimmlingen}}, \bibinfo {author} {\bibfnamefont {E.}~\bibnamefont
  {Eberlein}}, \bibinfo {author} {\bibfnamefont {T.}~\bibnamefont {Witzel}},
  \bibinfo {author} {\bibfnamefont {J.}~\bibnamefont {Cohen-Adad}}, \bibinfo
  {author} {\bibfnamefont {J.~A.}\ \bibnamefont {McNab}}, \bibinfo {author}
  {\bibfnamefont {B.}~\bibnamefont {Keil}}, \bibinfo {author} {\bibfnamefont
  {M.~D.}\ \bibnamefont {Tisdall}}, \bibinfo {author} {\bibfnamefont
  {P.}~\bibnamefont {Hoecht}}, \bibinfo {author} {\bibfnamefont
  {P.}~\bibnamefont {Dietz}}, \bibinfo {author} {\bibfnamefont {S.~F.}\
  \bibnamefont {Cauley}}, \bibinfo {author} {\bibfnamefont {V.}~\bibnamefont
  {Tountcheva}}, \bibinfo {author} {\bibfnamefont {V.}~\bibnamefont {Matschl}},
  \bibinfo {author} {\bibfnamefont {V.~H.}\ \bibnamefont {Lenz}}, \bibinfo
  {author} {\bibfnamefont {K.}~\bibnamefont {Heberlein}}, \bibinfo {author}
  {\bibfnamefont {A.}~\bibnamefont {Potthast}}, \bibinfo {author}
  {\bibfnamefont {H.}~\bibnamefont {Thein}}, \bibinfo {author} {\bibfnamefont
  {J.}~\bibnamefont {Van~Horn}}, \bibinfo {author} {\bibfnamefont
  {A.}~\bibnamefont {Toga}}, \bibinfo {author} {\bibfnamefont {F.}~\bibnamefont
  {Schmitt}}, \bibinfo {author} {\bibfnamefont {D.}~\bibnamefont {Lehne}},
  \bibinfo {author} {\bibfnamefont {B.~R.}\ \bibnamefont {Rosen}}, \bibinfo
  {author} {\bibfnamefont {V.}~\bibnamefont {Wedeen}}, \ and\ \bibinfo {author}
  {\bibfnamefont {L.~L.}\ \bibnamefont {Wald}},\ }\href {\doibase
  10.1016/j.neuroimage.2013.05.078} {\bibfield  {journal} {\bibinfo  {journal}
  {Neuroimage}\ }\bibinfo {series} {Mapping the {Connectome}},\ \textbf
  {\bibinfo {volume} {80}},\ \bibinfo {pages} {220} (\bibinfo {year}
  {2013})}\BibitemShut {NoStop}%
\end{thebibliography}%

\end{document}